\documentclass[10pt,twocolumn]{article}

\usepackage{scicite}

\usepackage{times}

\usepackage{graphicx}
\usepackage{amssymb}
\usepackage{multibib}
\usepackage{color}
\usepackage{abstract}

\newcommand{\vc}[1]{\mbox{\boldmath{$#1$}}}
\newcommand{\dpa}[0]{\partial}

\DeclareMathSymbol{\varOmega}{\mathord}{letters}{"0A}
\DeclareMathSymbol{\varPhi}{\mathord}{letters}{"08}
\DeclareMathSymbol{\varSigma}{\mathord}{letters}{"06}
\DeclareMathSymbol{\varPsi}{\mathord}{letters}{"09}
\DeclareMathSymbol{\varGamma}{\mathord}{letters}{"00}

\newcounter{lastnote}

\date{}
\title{Growth of Asteroids, Planetary Embryos and Kuiper Belt Objects\\ by
Chondrule Accretion}

\begin{document} 

\twocolumn[
\maketitle 
\begin{center}\vspace{-1.5cm}
Anders Johansen$^{1}$, Mordecai-Mark Mac Low$^{2}$, Pedro Lacerda$^{3}$, Martin
Bizzarro$^{4}$
\end{center}
{\footnotesize
$^{1}$Lund Observatory, Department of Astronomy and Theoretical Physics, Lund
University, Box 43, 22100 Lund, Sweden\\
$^{2}$Department of Astrophysics, American Museum of Natural History, 79th
Street at Central Park West, New York, NY 10024-5192, USA\\
$^{3}$Max Planck Institute for Solar System Research, Justus-von-Liebig-Weg 3,
37077 G\"ottingen, Germany \\
$^{4}$Centre for Star and Planet Formation and Natural History Museum of
Denmark, University of Copenhagen, \O ster Voldgade 5-7, 1350 Copenhagen,
Denmark}
\begin{onecolabstract}
Chondrules are millimeter-sized spherules that dominate primitive meteorites
(chondrites) originating from the asteroid belt. The incorporation of
chondrules into asteroidal bodies must be an important step in planet
formation, but the mechanism is not understood. We show that the main growth of
asteroids can result from gas-drag-assisted accretion of chondrules. The
largest planetesimals of a population with a characteristic radius of 100 km
undergo run-away accretion of chondrules within $\sim$3 Myr, forming planetary
embryos up to Mars sizes along with smaller asteroids whose size distribution
matches that of main belt asteroids. The aerodynamical accretion leads to
size-sorting of chondrules consistent with chondrites. Accretion of mm-sized
chondrules and ice particles drives the growth of planetesimals beyond the ice
line as well, but the growth time increases above the disk life time outside of
25 AU. The contribution of direct planetesimal accretion to the growth of both
asteroids and Kuiper belt objects is minor. In contrast, planetesimal accretion
and chondrule accretion play more equal roles for the formation of Moon-sized
embryos in the terrestrial planet formation region.  These embryos are isolated
from each other and accrete planetesimals only at a low rate. However, the
continued accretion of chondrules destabilizes the oligarchic configuration and
leads to the formation of Mars-sized embryos and terrestrial planets by a
combination of direct chondrule accretion and giant impacts.
\end{onecolabstract}
]

\noindent {\bf \large Introduction}
\vspace{0.2cm}

The formation of planetesimals and planetary embryos is an important step
towards the assembly of planetary systems
\cite{Johansen+etal2014,Raymond+etal2014}. The asteroid belt contains
planetesimals left over from the formation of the solar system and thus
provides a record of the early stages of planet formation.  Chondrite
meteorites are fragments of asteroids that did not undergo large-scale melting
and differentiation. Their constituent particles allow us to study the first
stage of planet formation, where micrometer-sized dust grains grew to
asteroid-scale planetesimals. The dominant mass fraction of most chondrite
meteorites are chondrules, millimeter-sized glassy spherules formed by
transient heating events in the protoplanetary disk \cite{Krot+etal2003}.
The small matrix grains that  fill the space between these chondrules
likely entered the meteorite parent bodies as accretionary rims on the
chondrules \cite{Metzler+etal1992}.

Uranium-corrected Pb-Pb dates of chondrules show crystallisation ages ranging
from the earliest epoch of the solar protoplanetary disk -- defined by the
condensation of calcium-aluminium-rich inclusions (CAIs) $4567.30\pm0.16$ Myr
ago -- to approximately 3 Myr later \cite{Connelly+etal2012}. Moreover,
chondrules from individual chondrites show variability in their
$^{54}$Cr/$^{52}$Cr and $^{50}$Ti/$^{47}$Ti ratios
\cite{Trinquier+etal2009,Connelly+etal2012} that track genetic relationships
between early-formed solids and their respective reservoirs.  Collectively,
these observations indicate that chondrules from individual chondrite groups
formed in different regions of the protoplanetary disk and were subsequently
transported to the accretion regions of their respective parent bodies. Thus,
the region of asteroid formation must have been dominated by chondrules over
time-scales comparable to the observed lifetimes of protoplanetary disks around
young stars \cite{Haisch+etal2001}.

Meteorites provide a direct view of the primitive, chondritic material from
which the planetesimals in the asteroid belt formed. In contrast, planetesimals
originally located in the formation region of terrestrial planets accreted into
planetary bodies and hence little evidence exists with respect to the material
from which those planetesimals formed. However, the abundance of chondrules in
the asteroid belt suggests that chondrules could have been widespread in the
terrestrial planet formation region as well. Indeed, some chondrules record
$^{54}$Cr/$^{52}$Cr and $^{50}$Ti/$^{47}$Ti ratios indicating formation in the
accretion regions of Earth and Mars
\cite{Trinquier+etal2009,Connelly+etal2012}. Hence, understanding the role of
chondrules in the formation of asteroids can provide critical insights into how
the terrestrial planets formed closer to the Sun.

The mechanism by which large amounts of chondrules were incorporated into
asteroids is poorly understood. Particles of chondrule sizes can concentrate
near the smallest scales of the turbulent gas\cite{Cuzzi+etal2001}.  The most
extreme concentration events have been proposed to lead to asteroid formation
by gravitational contraction of such intermittent chondrule
clouds\cite{Cuzzi+etal2008}. However, it is still debated whether such high
concentrations actually occur \cite{Pan+etal2011}. The streaming
instability\cite{YoudinGoodman2005} is an alternative mechanism that
concentrates larger particles\cite{Johansen+etal2007,Johansen+etal2009}, of
typical sizes from 10 cm to 1 m when applied to the asteroid belt, due to an
aerodynamical effect where particles pile up in dense filaments, which can
reach densities of more than 1,000 times the local gas
density\cite{Johansen+etal2012}. The streaming instability nevertheless fails
to concentrate chondrule-sized particles whose motion is strongly damped by gas
drag\cite{BaiStone2010}.
\begin{figure}[!t]
  \begin{center}
    \includegraphics[width=0.9\linewidth]{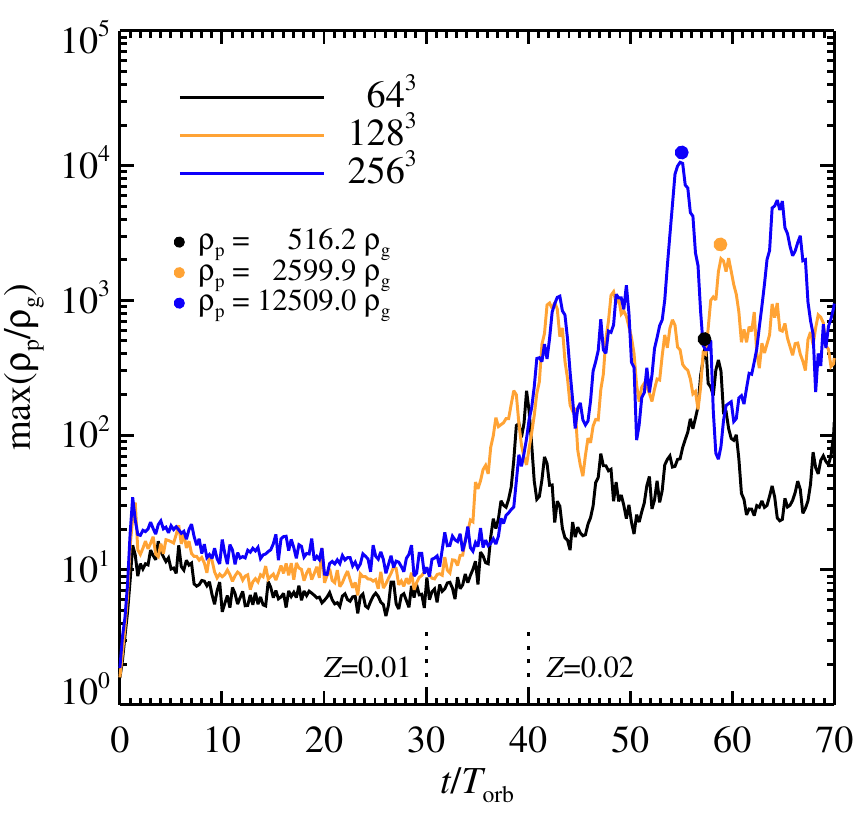}
  \end{center}
  \caption{The maximum particle density versus time for streaming instability
  simulations without self-gravity at resolutions $64^3$ (black line),
  $128^3$ (yellow line) and $256^3$ (blue line). The particle density is
  measured in units of the mid-plane gas density $\rho_{\rm g}$ and the time in
  units of the orbital period $T_{\rm orb}$. The first 30 orbits are run with
  a solids-to-gas ratio of $Z=0.01$, with only modest overdensities seen in
  the particle density.  Half of the gas is then removed over the next 10
  orbits, triggering strong particle concentration, of up to 10,000 times the
  local gas density at the highest resolution. Doubling the resolution leads to
  a more than a quadrupling of the maximum particle density.}
  \label{f:rhopmax_t}
\end{figure}

\vspace{0.2cm}
\noindent {\bf \large Results}
\vspace{0.2cm}

Here we report the results of a model where asteroids and planetary embryos
grow by accretion of chondrules onto planetesimal seeds formed by the streaming
instability.  Particles of large enough size to concentrate by streaming
instabilities could have been present in the early stages of the protoplanetary
disk when planetesimals formed. Such particles include macrochondrules
\cite{WeyrauchBischoff2012}, chondrule aggregates \cite{Ormel+etal2008} and
ice-rich pebbles. All of these drift rapidly towards the Sun because of gas
drag. Therefore we assume that cm-sized particles were only present in the
earliest stages of the protoplanetary disk, whereas smaller chondrules existed
during most of the disk's life-time. Chondrule-sized particles may have been
able to remain in the asteroid belt due to stellar outflows \cite{Shu+etal1996}
or advection with the outwards moving gas in the mid-plane layer of
sedimented particles \cite{BaiStone2010}. We discuss the conditions for
planetesimal formation in the asteroid belt, particularly the formation of
dm-sized particles, further in the Supplementary Text.
\begin{figure}[!t]
  \begin{center}
    \includegraphics[width=0.9\linewidth]{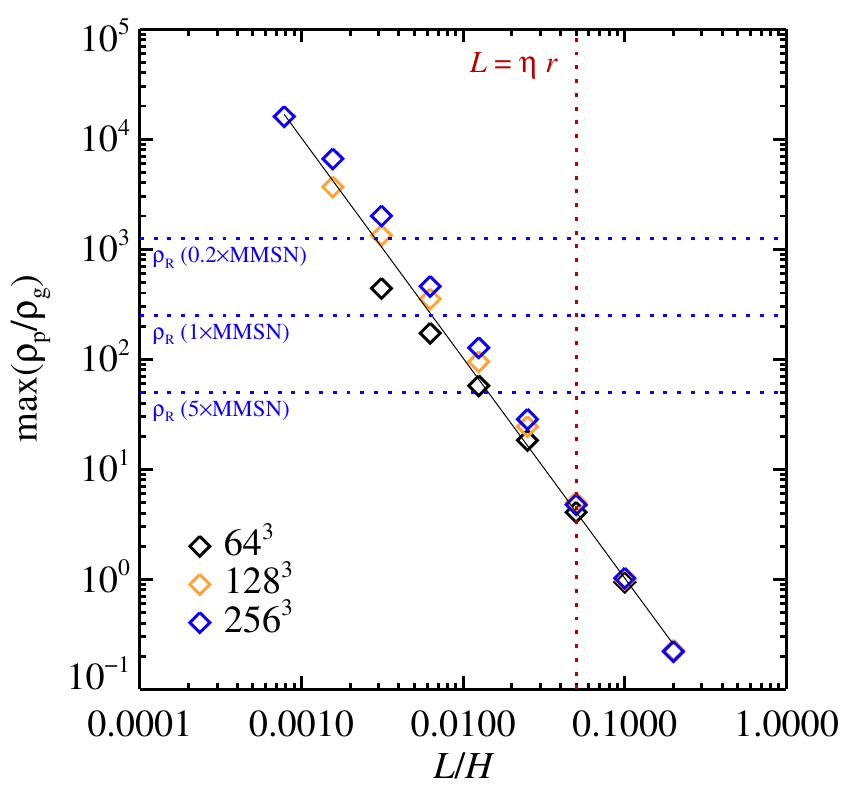}
  \end{center}
  \caption{The maximum particle density versus the length scale (measured
  in scale-heights $H$). We have taken spheres with diameters from one grid
  cell up to the largest scale of the simulation and noted the maximum value of
  the density over all simulation snapshots. The results are relatively
  converged on each scale. The increase in maximum particle density with
  increasing resolution is an effect of resolving ever smaller-scale
  filaments. Blue dotted lines mark the Roche density at three values of the
  gas column density at 2.5 AU compared to the Minimum Mass Solar Nebula
  (MMSN).  The red dotted line indicates the characteristic length scale of the
  streaming instability. The black line shows a power law of slope $-2$, which
  shows that the maximum density follows approximately the inverse square of
  the length scale.}
  \label{f:rhopmax_scale}
\end{figure}

\noindent {\bf Planetesimal formation simulations}

We use high-resolution models of planetesimal formation through streaming
instabilities to set the initial conditions for our model. The highest
resolution reached prior to this work in such simulations is $128^3$ grid cells
with $2.4$ million superparticles representing the solids
\cite{Johansen+etal2012}. We here report on simulations with up to $512^3$ grid
cells and $153.6$ million superparticles representing the solids, which is 64
times more resolution elements than in previous simulations. Details of the
simulation method are given in Materials and Methods.

We first perform a convergence test of streaming instability simulations with
no self-gravity between the particles. Figure \ref{f:rhopmax_t} shows the
maximum particle density versus time in these simulations. We run the initial
30 orbits with the full gas density. Here the maximum particle density is very
moderate and no strong concentrations occur. Between 30 and 40 orbits we remove
50\% of the gas, effectively increasing the solids-to-gas ratio from $Z=0.01$
to $Z=0.02$.  This increase in metallicity triggers strong particle
concentration through the streaming instability, of up to 10,000 times the
local gas density. The maximum particle density increases approximately
quadratically with the inverse size of the resolution element $\delta x$. The
maximum particle density on different scales is shown in Figure
\ref{f:rhopmax_scale}.  Here we have measured the maximum density in spheres of
diameters from $\delta x$ up to the full size of the simulation box. The
results are well converged from $64^3$ to $256^3$ at the scales that are
present at all resolutions. The smaller scales made accessible at higher
resolution yield increasingly higher densities. This is an effect of the
filamentary structure of the overdense particle filaments forming by the
streaming instability; higher resolution allows us to resolve finer structure
and hence higher densities. Smaller planetesimals can thus form as the
resolution is increased and increasingly smaller-scale filaments cross the
Roche density.
\begin{figure}[!t]
  \begin{center}
    \includegraphics[width=\linewidth]{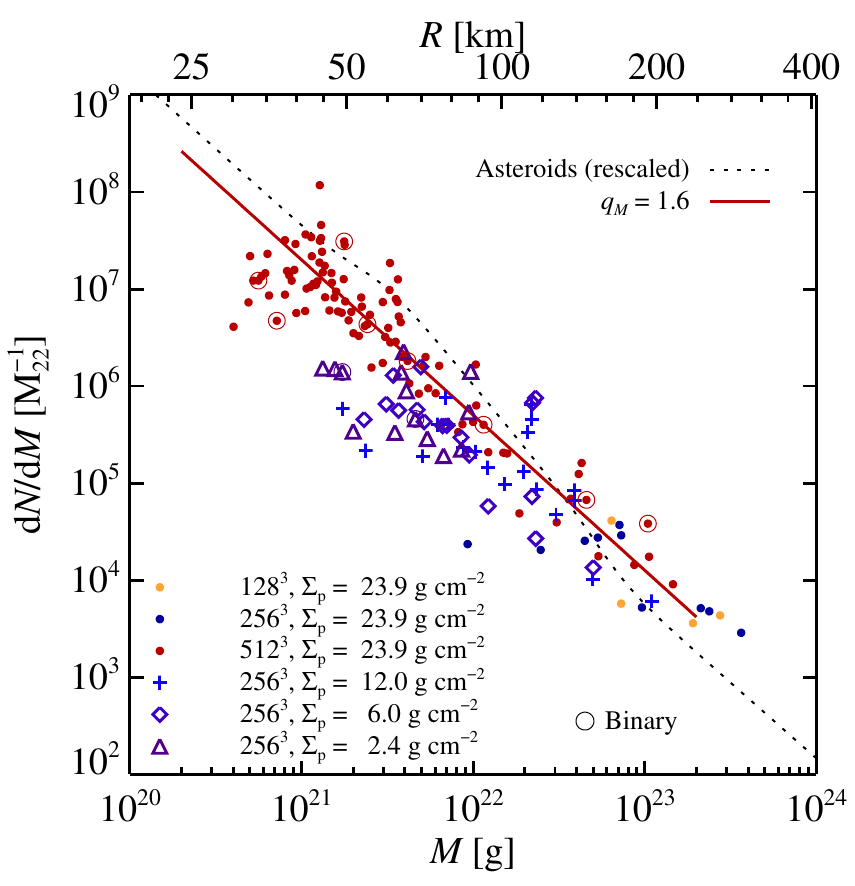}
  \end{center}
  \caption{Birth size distribution of planetesimals forming by the streaming
  instability in 25-cm-sized particles. The differential number of
  planetesimals (per $10^{22}$ g) is calculated with respect to the nearest
  size neighbours in the simulation. Yellow, blue and red circles indicate
  individual planetesimals forming in computer simulations at $128^3$, $256^3$
  and $512^3$ grid cells. The size distribution of the highest-resolution
  simulation is fitted well with a power law ${\rm d}N/{\rm d}M \propto
  M^{-q_M}$ (red line; the fit is based on the cumulative size distribution
  shown in
  Figure \ref{f:Ncumu_M_SI}, but does not include the exponential tapering).
  Simulations with lower values of the particle column density $\varSigma_{\rm
  p}$ yield successively smaller radii for the largest planetesimals, down to
  below 100 km in radius for a column density similar to the Minimum Mass Solar
  Nebula model ($\varSigma_{\rm p} = 4.3\,{\rm g\,cm^{-2}}$ at $r=2.5\,{\rm
  AU}$).  Binary planetesimals (marked with circles) appear only in the
  highest-resolution simulation, as binary survival requires sufficient
  resolution of the Hill radius. The differential size distribution of the
  asteroid belt (dashed line) shows characteristic bumps at 60 km and 200 km.
  The planetesimal birth sizes from the simulations are clearly not in
  agreement with main belt asteroids.}
  \label{f:size_distribution}
\end{figure}

In Figure \ref{f:size_distribution} we show size distributions of planetesimals
forming in streaming instability simulations that include the self-gravity of
the particles. The planetesimal size distribution is dominated in number by
small planetesimals, but most of the mass is in the few largest bodies. We find
that progressively smaller planetesimals form, alongside the large
planetesimals, at higher resolution. The largest planetesimals, which contain
the dominant mass of the population, decrease to approximately 100 km in radius
for the column density of the Minimum Mass Solar Nebula at the location of the
asteroid belt \cite{Hayashi1981}. The size distribution of asteroids above 60
km in radius has been shown to retain its primordial shape, since the depletion
of the asteroid belt happens in a size-independent fashion for those large
bodies\cite{Bottke+etal2005,Morbidelli+etal2009}. The differential size
distribution resulting from streaming instabilities disagrees with the observed
size distribution of the asteroids (dashed line in Figure
\ref{f:size_distribution}).

\noindent {\bf Chondrule accretion on asteroids}

We proceed to consider the subsequent evolution of the size distribution as the
newly formed planetesimals accrete chondrules present in the gaseous
surroundings. We assume that planetesimals are born within an ocean of small
chondrules that do not directly participate in planetesimal formation.
Accretion of these chondrules by the planetesimals over the following millions
of years can lead to significant mass growth. We use initial size distributions
inspired by the streaming instability simulations. However, chondrule accretion
is not limited to this model; it would be efficient in the context of any
planetesimal formation mechanism that produces planetesimals with
characteristic sizes of 100 km\cite{Cuzzi+etal2008,Cuzzi+etal2010}.
\begin{figure*}[!t]
  \begin{center}
    \includegraphics[width=0.8\linewidth]{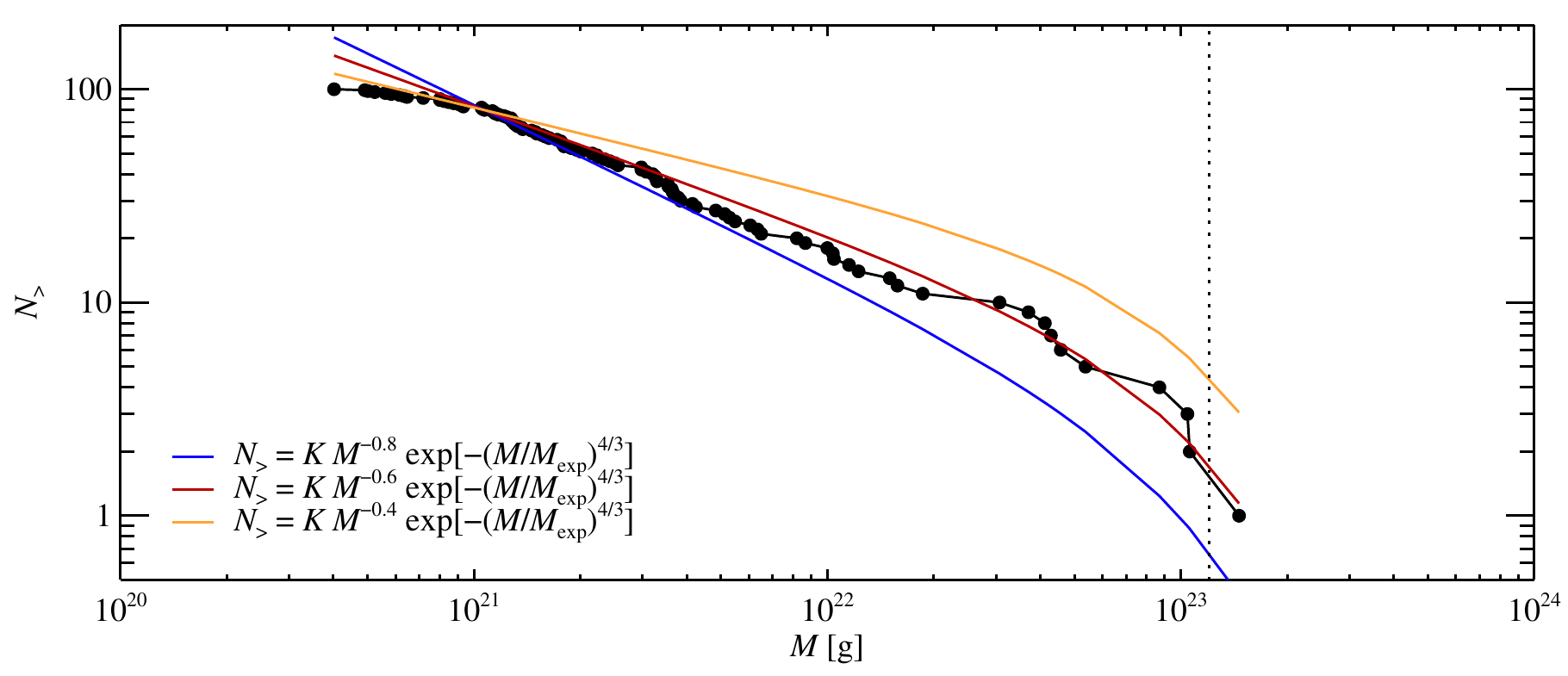}
  \end{center}
  \caption{Cumulative birth size distribution of planetesimals forming by the
  streaming instability in our highest-resolution simulation ($512^3$). The
  cumulative size distribution is fitted with an exponentially tapered power
  law $N_> \propto M^{-0.6} \exp[-(M/M_{\rm exp})^{(4/3)}]$ (red line), with
  exponential tapering at $M_{\rm exp}=1.2 \times 10^{23}$ g (dotted
  line). Shallower or steeper power laws yield poorer fits to the populations
  of small and large planetesimals, respectively. We choose to fit the
  cumulative size distribution rather than the differential size distribution
  to avoid the noise inherently present in the latter. The fit can be
  translated to ${\rm d}N/{\rm d}M \propto M^{-1.6} \exp[-(M/M_{\rm
  exp})^{4/3}]$ (the power law part of this fit is indicated in Figure
  \ref{f:size_distribution}) as well as to ${\rm d}N/{\rm d}R \propto R^{-2.8}
  \exp[-(R/R_{\rm exp})^4]$ in the differential size distribution per unit
  radius.}
  \label{f:Ncumu_M_SI}
\end{figure*}

The gas in the protoplanetary disk is slightly pressure-supported in the radial
direction and hence moves at a sub-Keplerian speed, with $\Delta v$ marking the
difference between the Keplerian speed and the actual gas speed. The speed
difference is approximately $53$ m/s in the optically thin Minimum Mass Solar
Nebula model \cite{Hayashi1981}. The Bondi radius of a planetesimal with
radius $R$ and internal density
$\rho_\bullet$,
\begin{equation}
  \frac{R_{\rm B}}{R} = 3.5 \left( \frac{R}{\rm 100\,km} \right)^2 \left(
  \frac{\Delta v}{\rm 53\,m/s} \right)^{-2} \left( \frac{\rho_\bullet}{\rm
  3.5\,g/cm^3} \right) \, ,
\end{equation}
measures the gravitational deflection radius of the planetesimal. Chondrules
embedded in the sub-Keplerian gas flow are accreted from the Bondi
radius\cite{OrmelKlahr2010,LambrechtsJohansen2012} when their friction time
$t_{\rm f}$ is comparable to the time to drift azimuthally across the Bondi
radius, $t_{\rm B}=R_{\rm B}/\Delta v$. An additional turbulent component
to the gas motion can be ignored since this is expected to be much slower than
the sub-Keplerian speed in the asteroid formation region. Peak accretion rates
are obtained for $t_{\rm f}/t_{\rm B}$ in the range from 0.5 to 10
\cite{LambrechtsJohansen2012}. This happens at the orbital distance of the
asteroid belt for particle sizes in the interval
\begin{equation}
  a = [0.08,1.6]\,{\rm mm}\, \left( \frac{R}{100\,{\rm km}} \right)^3 \left(
  \frac{\Delta v}{53\,{\rm m/s}} \right)^{-3} f_{\rm gas} \, .
  \label{eq:aopt}
\end{equation}
Here $f_{\rm gas}$ represents the ratio of the actual gas column density to
that of the Minimum Mass Solar Nebula \cite{Hayashi1981}.  Chondrule
accretion happens at a rate proportional to $R_{\rm B}^2$ (or $\dot{M} \propto
R^6$). This run-away accretion can drive a very steep differential size
distribution similar to what is observed for large asteroids in the asteroid
belt.
\begin{figure*}[!t]
  \begin{center}
    \includegraphics[width=0.472\linewidth]{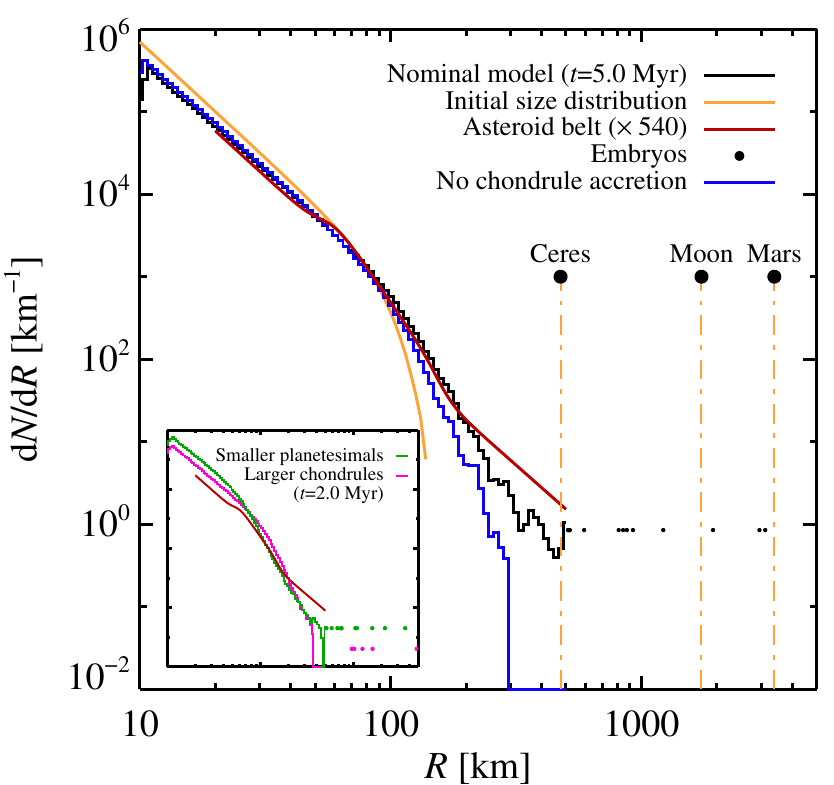}
    \includegraphics[width=0.508\linewidth]{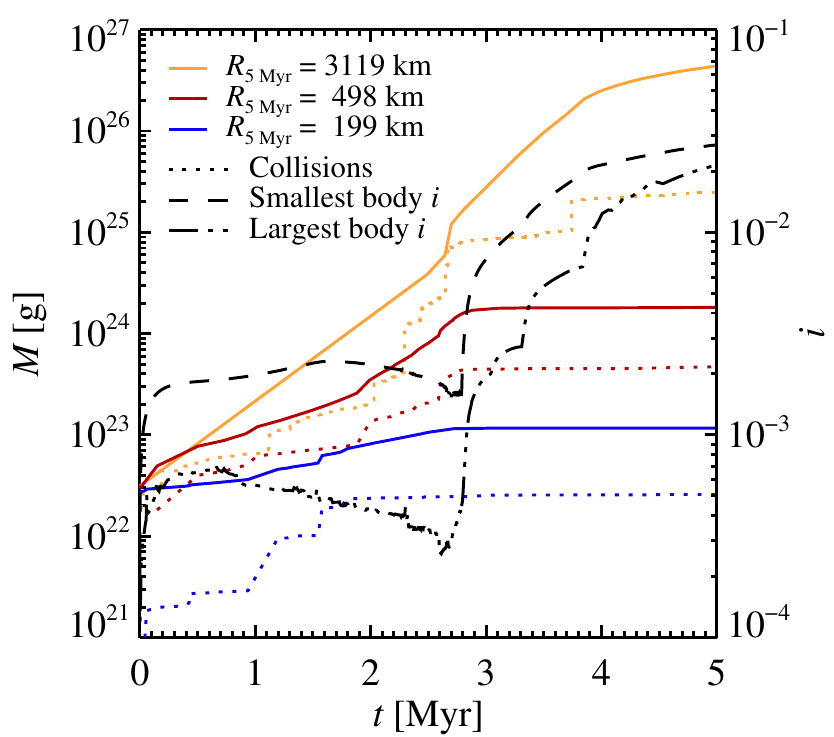}
  \end{center}
  \caption{The size distribution of asteroids and embryos after accreting
  chondrules for 5 Myr (left panel) and selected masses and inclinations as a
  function of time (right panel). The nominal model (black line) matches
  closely the steep size distribution of main belt asteroids (red line,
  representing the current asteroid belt multiplied by a depletion factor of
  540) from 60 km to 200 km in radius. The size distribution becomes shallower
  above 200 km; this is also seen in the asteroid belt, although the
  simulations underproduce Ceres-sized planetesimals by a factor of
  approximately 2--3. A simulation with no chondrules (blue line) produces no
  asteroids larger than 300 km.  Inclusion of chondrules up to cm sizes
  (pink line in insert) gives a much too low production of Ceres-sized
  asteroids, while setting the exponential cut-off radius of planetesimals to
  50 km (green line) leads to a poorer match to the bump at 60 km. The right
  plot shows that the formation of the first embryos after 2.5 Myr quenches
  chondrule accretion on the smaller asteroids by exciting their inclinations
  $i$ (right axis). The dotted lines indicate the mass contribution from
  planetesimal-planetesimal collisions.  Asteroids and embryos larger than 200
  km in radius owe at least 2/3 of their mass to chondrule accretion.}
  \label{f:planetesimal_variation}
\end{figure*}
\begin{figure}[!t]
  \begin{center}
    \includegraphics[width=\linewidth]{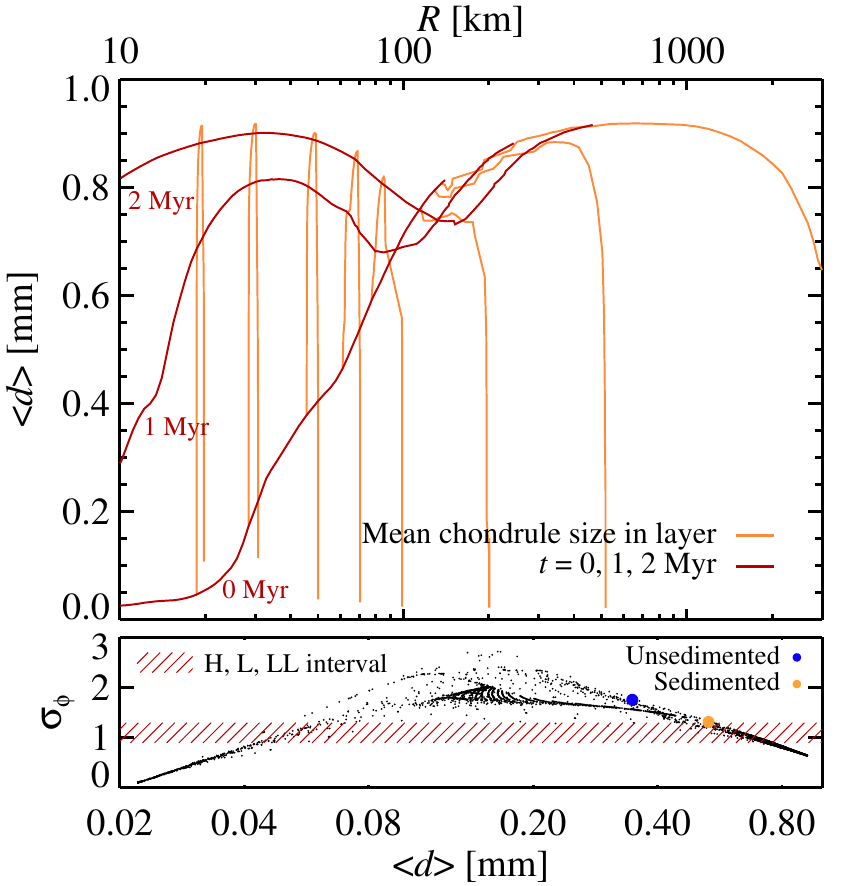}
  \end{center}
  \caption{Mean chondrule sizes ($\langle d \rangle$, upper panel) as a
  function of layer radius $R$ and size-distribution width ($\sigma_\phi$,
  lower panel) as a function of mean chondrule size. Yellow lines in the upper
  panel indicate the chondrule size evolution in individual asteroids and
  embryos, while red lines indicate mean accreted chondrule sizes at different
  times. The accreted chondrule size increases approximately linearly with
  planetesimal size at $t=0$ Myr. Asteroids stirred by the growing embryos over
  the next 2 Myr accrete increasingly larger chondrules, as asteroid velocities
  align with the sub-Keplerian chondrule flow at aphelion. Finally, asteroids
  accrete surface layers of mainly very small chondrules, down to below 0.1 mm
  in diameter, at late times when their large inclinations decouple the
  asteroid orbits from the large chondrules in the mid-plane. The width of the
  chondrule size distribution in the lower panel is given in terms of
  $\sigma_\phi$, the base-2-logarithmic half-width of the cumulative mass
  distribution of chondrules ($\sigma_\phi=1$ implies that 2/3 of the
  chondrules have diameters within a factor $2^1=2$ from the mean). Dots
  indicate chondrule layers inside asteroids and embryos. Chondrules are
  aerodynamically sorted by the accretion process, to values of $\sigma_\phi$
  in agreement with the chondrules found in ordinary chondrites (hashed
  region). Unfiltered accretion from the background size distribution of
  chondrules (blue dot: size distribution of un-sedimented particles;
  yellow dot: size distribution in the mid-plane) yields specific pairs of
  $\langle d \rangle$ and $\sigma_\phi$ that are not consistent with
  constraints from ordinary chondrites.}
  \label{f:meand_meanphi_R}
\end{figure}

We solve numerically for the temporal evolution of the sizes and orbits of
planetesimals accreting chondrules and colliding with each other in an annulus
extending from 2.4 to 2.6 AU (see Materials and Methods for details).
The planetesimals have initial radii from
10 to 150 km, distributed in size along an exponentially tapered power law
based on the cumulative size distribution of planetesimals forming in the
streaming instability simulations (see Figure \ref{f:Ncumu_M_SI}), ${\rm
d}N/{\rm d}R \propto R^{-2.8} \exp[-(R/R_{\rm exp})^4]$, with exponential
tapering at $R_{\rm exp}=100$ km and a total mass of 0.04 Earth masses.
Chondrules have diameters from 0.02 mm to 1.6 mm, in broad agreement with
observed chondrule sizes\cite{Krot+etal2003}. The growth of particles and the
formation of chondrule precursors is discussed in the Supplementary Material.
We place 0.2 Earth masses of chondrules initially in the annulus while another
0.2 Earth masses is created continuously over a time span of 3 Myr. The
chondrules are given a Gaussian density distribution around the mid-plane, with
the scale-height set through a nominal value of the turbulent viscosity of
$\alpha=10^{-4}$. The gas in the protoplanetary disk is depleted exponentially
on a time-scale of 3 Myr \cite{Haisch+etal2001}.

We show in the left panel of Figure \ref{f:planetesimal_variation} the size
distribution after 5 Myr of accreting chondrules. The steep drop in the initial
number of planetesimals larger than 100 km in radius has become much shallower
because of chondrule accretion, forming a characteristic bump at 60-70 km
radius. The size distribution of smaller asteroids in today's belt was likely
filled later with fragments from collisional grinding\cite{Bottke+etal2005}.
The steep size distribution of larger asteroids ends at around 200 km as even
larger asteroids would need to accrete chondrules larger than millimeter in
size (see equation 2). At this point the largest asteroids enter a more
democratic growth phase where the accretional cross sections are reduced by gas
friction within the Bondi radius\cite{OrmelKlahr2010,LambrechtsJohansen2012}.
The agreement with the observed size distribution of asteroids is excellent in
the nominal model, except in the range around Ceres mass where it predicts too
few objects by a factor of approximately 2--3. Changing the model parameters
(using larger chondrules or smaller planetesimal birth sizes) yields poorer
agreement with observed asteroid sizes. This implies that the observed
size distribution of asteroids is directly determined by the size distribution
of the chondrules and the birth sizes of the planetesimals.

The right panel of Figure \ref{f:planetesimal_variation} shows that the major
growth phase of asteroids with final radii ranging from 200 km to 500 km occurs
after 2.5 Myr.  Beyond this time the largest planetesimals in the population
grow to become planetary embryos with sizes similar to the Moon and Mars. These
growing embryos excite the inclinations of the smaller asteroids to
$i\sim0.01$, which disconnects the asteroids from the chondrules in the
mid-plane layer, quenching their accretion of further chondrules. The
beginning of embryo formation terminates the accretion of asteroids after 3
Myr, thus defining the final sizes of the asteroid belt population.  The
absence of such planetary embryos in today's asteroid belt may reflect a later
depletion by gravitational perturbations from Jupiter
\cite{Wetherill1992,MintonMalhotra2009,Walsh+etal2011}. Depletion of the
asteroid belt is discussed further in the Supplementary Text. An important
implication of chondrule accretion is that accretion of other planetesimals
contributes only a minor fraction to the final masses of large asteroids and
embryos (right panel of Figure \ref{f:planetesimal_variation}).

\noindent {\bf Size sorting of chondrules}

Chondrules in chondrites appear strongly size-sorted
\cite{Dodd1976,Cuzzi+etal2001}, with the average chondrule diameter varying
among the ordinary chondrites \cite{Dodd1976} from 0.3 mm (H chondrites) up to
0.5 mm (L and LL chondrites). Carbonaceous chondrites exhibit a larger range in
chondrule sizes, from 0.1 to 1 mm.  In Figure \ref{f:meand_meanphi_R} we show
that the mean diameter of accreted chondrules in our model lies in a range
similar to that observed for chondrites. The size distribution of accreted
chondrules is very narrow, also in agreement with the observed size
distribution in the ordinary chondrites\cite{Dodd1976}. The chondrule accretion
process, which leads to aerodynamical sorting of chondrules, may thus represent
the underlying physical mechanism for the size sorting of chondrules observed
in chondrite meteorites.  Although aerodynamical sorting has been previously
proposed to arise from the gas flow around asteroids \cite{Whipple1971}, our
simulations show that accretion from the full Bondi radius is much more
efficient in achieving aerodynamical sorting of chondrules.

\noindent {\bf Terrestrial planet formation with chondrules}

\begin{figure*}[!t]
  \begin{center}
    \includegraphics[width=0.9\linewidth]{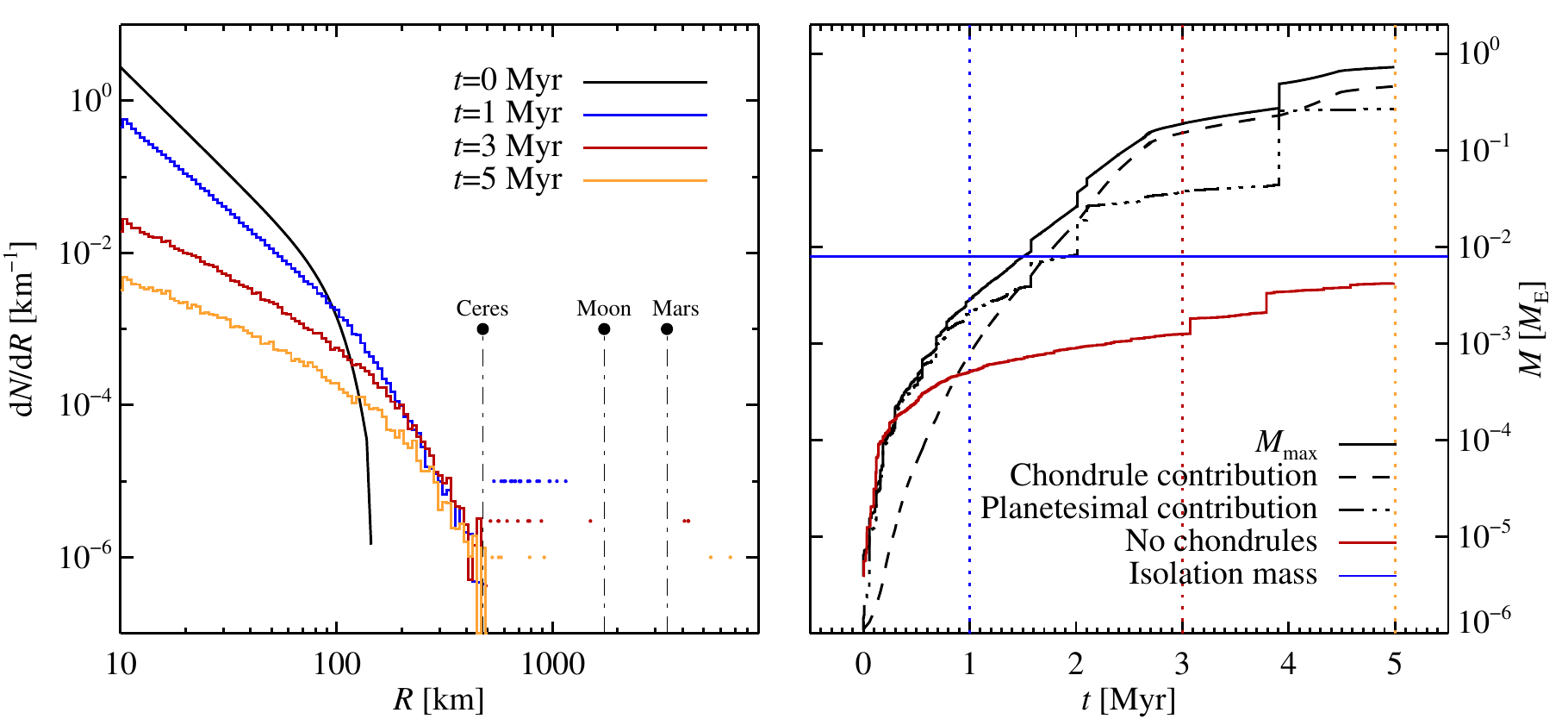}
  \end{center}
  \caption{Growth of embryos and terrestrial planets at 1 AU. The left panel
  shows the size distribution at four different times and the right panel shows
  the mass of the most massive body in the simulation as a function of time.
  The growth up to 1000-km-sized embryos is mainly driven by planetesimal
  accretion, since chondrule-sized pebbles are tightly coupled to the gas and
  hence hard to accrete. However, chondrule accretion gradually comes to
  dominate the accretion as the embryos grow. The largest body reaches Mars
  size after 3 Myr, with more than 90\% contribution from chondrule accretion.
  A giant impact occurs just before 4 Myr, where the largest body accretes the
  third largest body in the population. The continued accretion of chondrules
  leads to the formation of an Earth-mass body after 5 Myr. A simulation with
  no chondrules evolves very differently: a large number of embryos form with
  masses just below the isolation mass of $M_{\rm iso}\approx0.01M_{\rm E}$.}
  \label{f:dNdR_R_Rmax_t_1AU}
\end{figure*}

Our identification of chondrule accretion as driving planetesimal growth in the
asteroid belt implies that chondrules could play an important role for
terrestrial planet formation as well. To test this hypothesis we have performed
a numerical integration of the evolution of planetesimal orbits and sizes at 1
AU (Figure \ref{f:dNdR_R_Rmax_t_1AU}). The left panel shows the size
distribution of the planetesimals at 0, 1, 3 and 5 Myr. Growth to
super-Ceres-sized planetesimals is rapid and happens within 1 Myr. This growth
is driven mainly by planetesimal accretion, in stark contrast to the situation
at 2.5 AU. This is seen in the right panel of Figure \ref{f:dNdR_R_Rmax_t_1AU}
where the mass of the largest body is shown as a function of time (full line)
together with the mass contribution from chondrules (dashed line) and from
planetesimals (dash-dotted line). However, chondrule accretion becomes dominant
after 1.5 Myr and drives the further growth up to Mars-mass embryos after 4
Myr.  The largest body experiences a giant impact after 4 Myr, after which it
continues to grow by chondrule accretion towards 0.9 Earth masses.

Chondrule accretion can thus promote the growth of the largest embryos from
Moon masses towards Mars masses and finally Earth masses. The dominance of
planetesimal accretion in the initial growth towards Moon masses occurs because
chondrules couple tightly to the gas at 1 AU. Here the gas density is more than
a factor 10 higher than at 2.5 AU. This prevents sedimentation, such that all
chondrule sizes are well mixed with the gas, and it reduces the cross section
for accreting chondrules since tightly coupled particles can not be accreted
from the full Bondi radius. This situation changes as gas dissipates
exponentially over 3 Myr, increasing the friction times and thus the accretion
efficiency of the chondrules.  Furthermore, the increasingly large embryos
obtain high accretion radii for chondrules and hence high chondrule accretion
rates, despite the relatively low degree of sedimentation of chondrules present
at 1 AU orbits.
\begin{figure*}
  \begin{center}
    \includegraphics[width=0.9\linewidth]{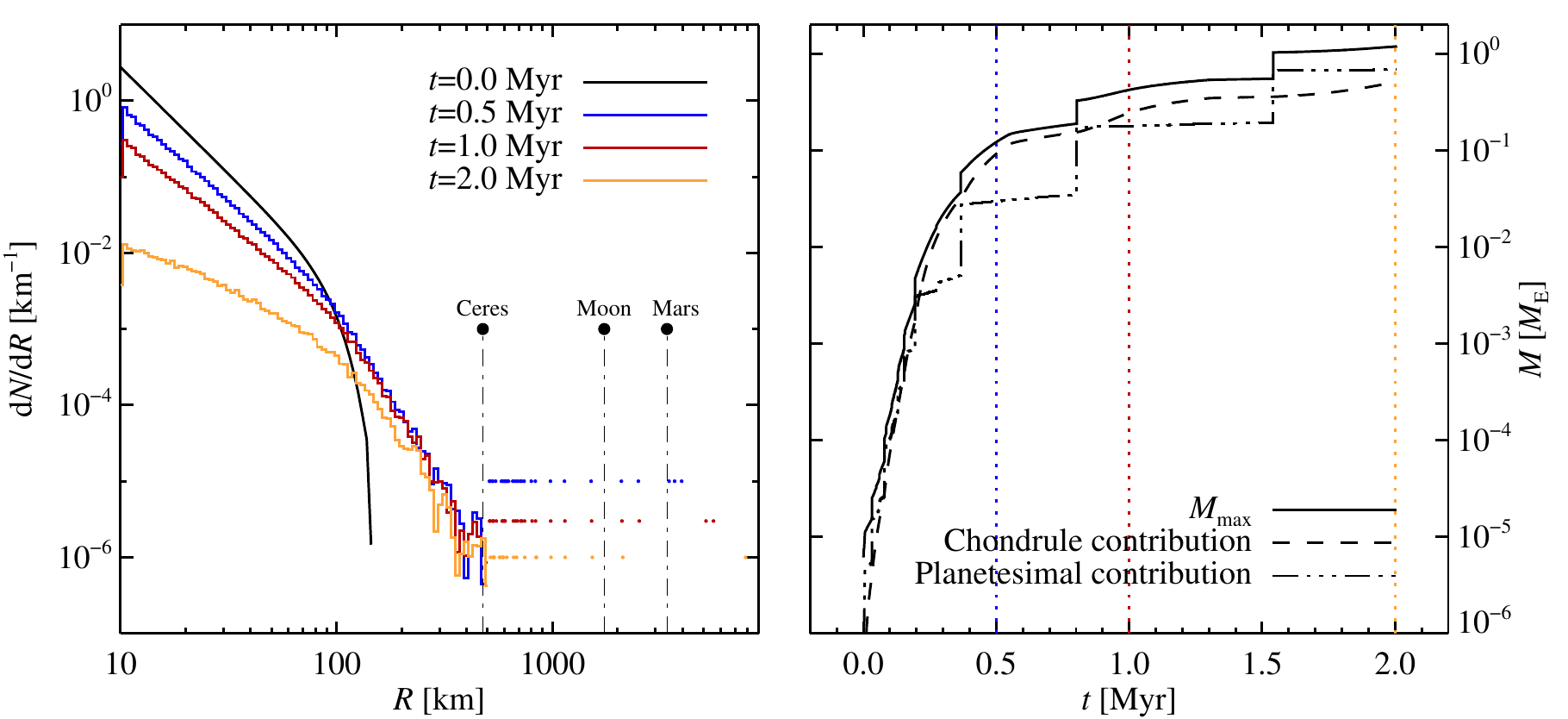}
  \end{center}
  \caption{Growth of embryos and terrestrial planets at 1 AU, with chondrule
  sizes up to 1 cm. The increase in chondrule size compared to the previous
  figure enhances the chondrule accretion rate substantially. The initial
  growth to 1000-km-sized embryos now has approximately equal contribution from
  planetesimal accretion and chondrule accretion. An Earth-mass terrestrial
  planet forms already after 2 Myr, driven by a combination of chondrule
  accretion and giant impacts.}
  \label{f:dNdR_R_Rmax_t_1AU_apmax1}
\end{figure*}

Chondrules also play a critical role in terrestrial planet formation in a
different way, namely by breaking the isolation mass configuration. Growth by
pure planetesimal accretion tends to end in oligarchic growth where the largest
embryos are isolated from each other by approximately 10 Hill radii
\cite{KokuboIda1998}. We have implemented the effect of this isolation tendency
by identifying the group of isolated bodies as those that can fit their
combined reach of 10 Hill radii into the annulus of 0.2 AU in diameter
\cite{Inaba+etal2001,Ohtsuki+etal2002}. Isolated bodies are not allowed to
accrete each other and only affect each other dynamically via distant viscous
stirring in the eccentricity. Inclination perturbations, as well as dynamical
friction in the eccentricity, are ignored between isolated bodies.

We mark the isolation mass of approximately 0.01 $M_{\rm E}$ in the right plot
of Figure \ref{f:dNdR_R_Rmax_t_1AU} (blue line). A simulation performed with no
chondrule accretion (red line) shows the expected growth of the largest embryos
to just below the isolation mass. The growth curve for pure planetesimal
accretion follows closely the growth curve for the full simulation until
approximately $10^{-4}$ Earth masses. At this point chondrule accretion becomes
significant and the two curves diverge. Importantly, chondrule accretion can
destabilise a set of isolated bodies, as their continued growth by chondrule
accretion drives the least massive of the isolated bodies out of isolation.
Hence the giant impacts experienced by the largest embryo between 1.5 and 4 Myr
are all driven by chondrule accretion, as these impacts happen beyond the
isolation mass.

The importance of chondrule accretion increases if chondrules in the
terrestrial planet formation region are larger, namely up to 1 cm
(macrochondrules). We show the results of expanding the size distribution of
chondrules up to 1 cm in Figure \ref{f:dNdR_R_Rmax_t_1AU_apmax1}. Here the
initial growth has equal contribution from chondrules and planetesimals. The
growth towards embryos and terrestrial planets is much more rapid than
in Figure \ref{f:dNdR_R_Rmax_t_1AU}; a planet with the mass of the Earth
already forms after 1.5 Myr.

The final size of embryos and planets forming by combined planetesimal
accretion and chondrule accretion depends on the exact timing of disk
dissipation. If the inner disk already dissipates after 3 Myr, then the
terrestrial planet formation region will be dominated by a number of Mars-sized
embryos. These bodies can go on to partake in a traditional terrestrial planet
formation scenario of gradual orbital perturbations and giant impacts over the
next 30--100 Myr \cite{OBrien+etal2006}. In contrast, later disk dissipation or
the presence of large chondrules in the terrestrial planet formation region
allows for the direct formation of planetary bodies within a few Myr. This is
far more rapid than traditional terrestrial planet formation scenarios that
do not consider chondrule accretion.

\noindent {\bf Pebble accretion in the Kuiper belt}

The size distribution of trans-Neptunian objects has a characteristic bump
around 50 km in radius \cite{SheppardTrujillo2010}, similar to main belt
asteroids, followed by a steep decline towards larger sizes
\cite{Bottke+etal2005}. We therefore explore here whether accretion of
chondrules and icy pebbles could be responsible for the observed size
distribution of Kuiper belt objects as well.

The asteroid belt displays a value of $q \approx 4.5$ in the size distribution
${\rm d}N/{\rm d}R \propto R^{-q}$ of asteroids larger than 60 km in radius,
while the hot population of the classical Kuiper belt and the Neptune trojans
have a much higher value of $q \approx 5 \ldots 6$ for large planetesimals
\cite{Fraser+etal2014}. The cold population of the classical Kuiper belt has an
even steeper decline, with $q \approx 8 \ldots 9$.  The planetesimals in the
cold population are believed to have formed in situ in the outermost regions of
the solar system \cite{Batygin+etal2011}, while the hot population and
scattered disk objects formed outside the birth location of Neptune and were
subsequently pushed outwards by Neptune's migration to its current orbit
\cite{Malhotra1995,Tsiganis+etal2005}. The planetesimals which formed between
the orbits of Jupiter and Neptune may today be represented by the C-type
asteroids, scattered to the asteroid belt during Jupiter's migration
\cite{Walsh+etal2011}. In order to probe planetesimal growth by chondrule
accretion (or icy particles of a similar size range) beyond the asteroid belt,
we therefore consider the two characteristic distances 10 AU and 25 AU.

\noindent {\it Planetesimal growth at 10 AU}

The region around 10 AU represents the conditions where the gas giants formed.
The formation of Jupiter and Saturn must have had an enormous effect on the
planetesimals in that region, with some ejected to the Oort cloud and others
potentially mixed into the outer asteroid belt during Jupiter's and Saturn's
migration \cite{Walsh+etal2011}. To test how the size distribution of
planetesimals evolves at 10 AU, we set up an annulus of width 0.2 AU. The
planetesimals are again given sizes from 10 to 150 km in radius, with a steep
exponential tapering above 100 km. The internal density of the planetesimals is
set to 2 g/cm$^3$, similar to that of Ceres, a typical representative of the
icy asteroids in the outer part of the main belt. The pebbles have radii from
0.01 to 1 mm. These pebbles could represent a mixture of chondrules transported
outwards from the inner solar system and icy pebbles formed beyond of the ice
line. Chondrules may also have formed in situ in the outer parts of the
protoplanetary disk. In fact, some chondrules present in carbonaceous
chondrites have $^{50}$Ti/$^{47}$Ti ratios comparable to those of the most
pristine chondrites, namely the CI carbonaceous chondrites
\cite{Trinquier+etal2009}. This suggests that at least a fraction of chondrules
in carbonaceous chondrites formed from pristine, thermally unprocessed
precursor material typical of the accretion regions of CI chondrites. Given
that these chondrites are believed to have accreted beyond the ice line
\cite{Walsh+etal2011}, chondrule formation appears to not have been limited to
the asteroidal belt, but also occurred in the outer solar system.

\begin{figure*}[!t]
  \begin{center}
    \includegraphics[width=0.9\linewidth]{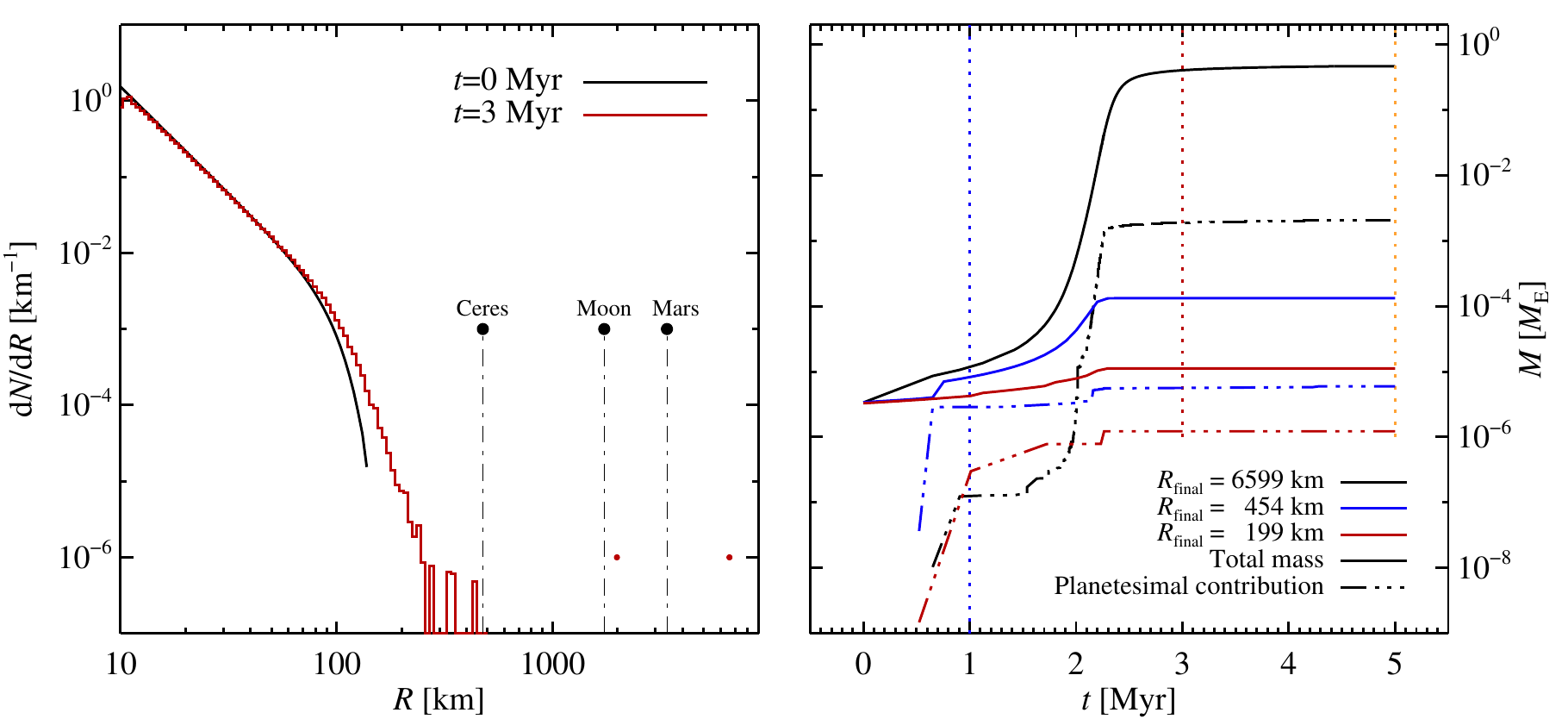}
  \end{center}
  \caption{Growth of icy planetesimals at 10 AU. The growth rate by pebble
  accretion is as high as in the asteroid belt, since the lower column density
  of pebbles is counter-acted by the increased sedimentation in the more dilute
  gas. The annulus of 0.2 AU width produces in the end an Earth-sized
  protoplanet and a single Moon-sized embryo. Pebble accretion overwhelmingly
  dominates the growth (right panel).  The icy protoplanet that forms has only
  a few parts in a thousand mass contribution from collisions. Large
  Ceres-sized planetesimals have a contribution from collisions of less than
  5\%, while the 200 km planetesimal owes about 1/10 of its growth from 130 km
  to planetesimal collisions. Note how the Ceres-sized planetesimal (blue line
  in the right panel) got a head start for efficient accretion of pebbles by
  experiencing a significant collision after 700,000 years.}
  \label{f:dNdR_R_R_t_10AU}
\end{figure*}

The results at 10 AU are shown in Figure \ref{f:dNdR_R_R_t_10AU}.
Planetesimals of Ceres size grow within approximately 2 Myr, followed by the
usual phase of embryo growth also seen in Figure
\ref{f:planetesimal_variation}. The largest embryo reaches a mass comparable to
the Earth's before 3 Myr. The mid-plane density of pebbles is much lower at 10
AU than at 2.5 AU, but this is more than counter-acted by increased
sedimentation of particles in the dilute gas. The size distribution is very
steep, much steeper than in the asteroid belt, and resembles better the
steepness of the size distribution of the trans-Neptunian populations
\cite{Fraser+etal2014}. Finally, it is interesting to note that at these
orbital distances planetesimal-planetesimal collisions are even less important
than at 2.5 AU (right panel of Figure \ref{f:dNdR_R_R_t_10AU}). The Earth-mass
protoplanet has only a few parts in a thousand contribution from planetesimal
collisions.
\begin{figure*}[!t]
  \begin{center}
    \includegraphics[width=0.9\linewidth]{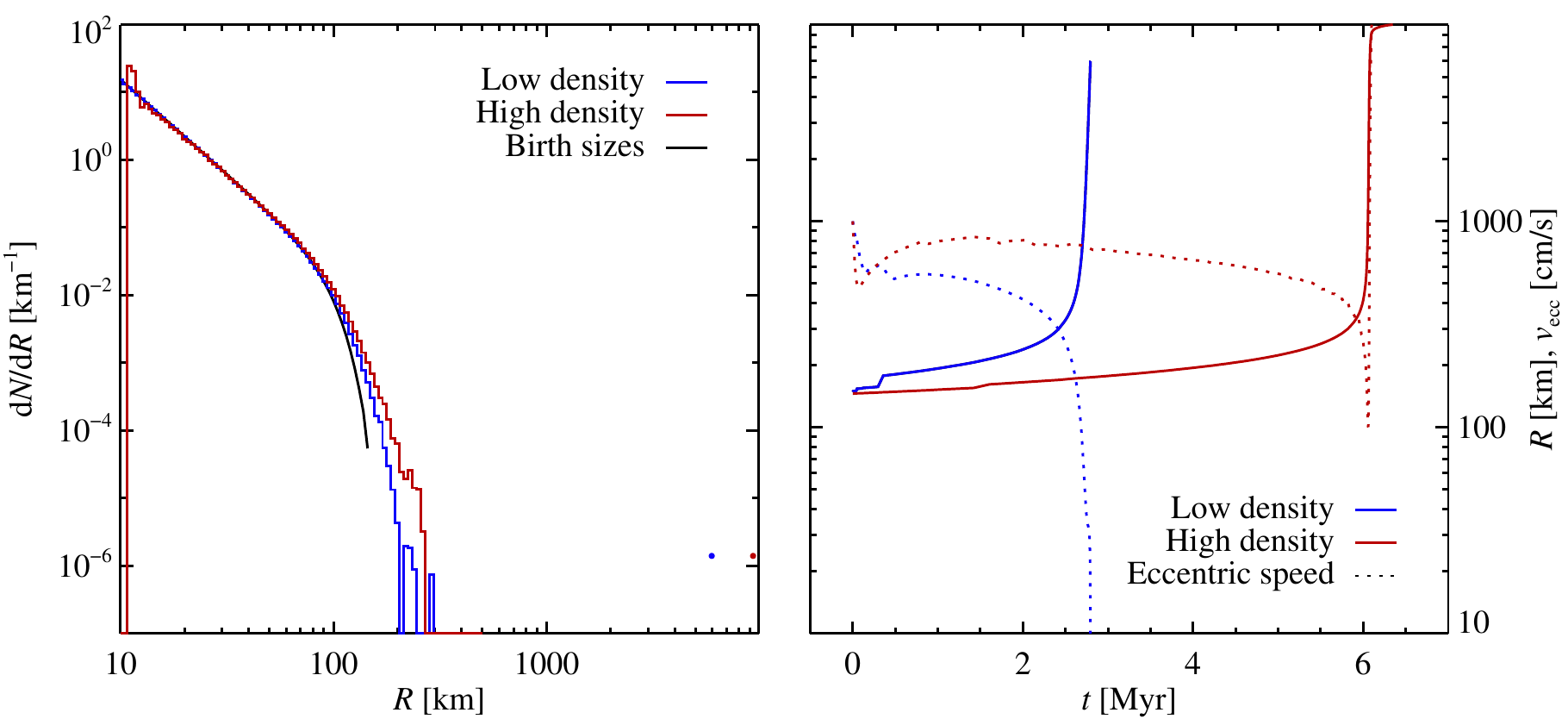}
  \end{center}
  \caption{Planetesimal growth at 25 AU. Two models are considered, a {\it
  low density model} where the internal density is set to $\rho_\bullet=0.5$
  g/cm$^3$, similar to comets and binary Kuiper belt objects, and a {\it high
  density model} where the internal density is set to $\rho_\bullet=2$
  g/cm$^3$, similar to the dwarf planet Ceres. The low density model has
  turbulent stirring $\alpha=10^{-6}$ while the high density model has
  $\alpha=10^{-5}$. Both models display ordered growth up to 300 km radii, with
  a steep size distribution beyond 100 km sizes. This is followed by a run-away
  growth of a single, massive body. The right panel shows the size of the
  largest body as a function of time as well as the speed relative to a
  circular orbit. The run-away growth is facilitated by a steep decline in the
  eccentricity of the orbit, as the high pebble accretion rate damps the
  eccentricity.}
  \label{f:dNdR_R_Rmax_t_25AU}
\end{figure*}

\noindent {\it Planetesimal growth at 25 AU}

In the outer regions of the protoplanetary disk we adopt a model where the
surface density profile is proportional to the inverse of the orbital radius.
This is in agreement with surface density profiles of observed protoplanetary
disks \cite{WilliamsCieza2011,Birnstiel+etal2012}. Hence we set the total
column density of particles at $r=25$ AU to $\varSigma_{\rm p}=0.54$ g/cm$^2$.
We set the temperature at 25 AU according to full radiative transfer models of
protoplanetary disks. This is important, since the optically thin assumption in
the Minimum Mass Solar Nebula model of Hayashi \cite{Hayashi1981} overestimates
the temperature in the outer disk. The growth rate of a planetesimal with mass
$M_{\rm p}$ by pebble accretion is given by
\begin{equation}
  \dot{M}_{\rm p} = \pi f_{\rm B}^2 \left[ \frac{G M_{\rm p}}{(\Delta v)^2}
  \right]^2 (\Delta v) \rho_{\rm p} \, .
\end{equation}
Here $f_{\rm B}$ denotes the ratio of the actual accretion radius to the Bondi
radius \cite{LambrechtsJohansen2012}, $\Delta v$ is the sub-Keplerian speed
difference and $\rho_{\rm p}$ is the mid-plane particle density. The
sub-Keplerian speed is given by
\begin{equation}
  \Delta v = - \frac{1}{2} \frac{H}{r} \frac{\dpa \ln P}{\dpa \ln r} c_{\rm s}
  \, .
\end{equation}
The speed difference scales with the sound speed as $c_{\rm s}^2$, since $H/r$
contains another $c_{\rm s}$. Hence the accretion rate scales with the sound
speed to the negative sixth power, and the sound speed must be set carefully to
yield realistic results. We set the sound speed at 25 AU through the aspect
ratio $H/r=0.05$, based on radiative transfer models of protoplanetary
accretion disks \cite{Bitsch+etal2014}. This is much smaller than the standard
value in the Minimum Mass Solar Nebula ($H/r=0.072$ at $r=25$ AU) and leads
to much higher accretion rates in the outer nebula.

We consider two models for planetesimal growth at 25 AU. First, a low density
model where the internal density of the planetesimals is set to 0.5 g/cm$^3$,
similar to comets and some binary Kuiper belt objects \cite{Brown2013}. Second,
a high density model that has instead an internal density of 2 g/cm$^3$, more
in agreement with the icy dwarf planets Pluto and Ceres. The low density model
experiences much lower pebble accretion rates because of the reduced gravity of
the planetesimals. We find that a turbulent stirring of $\alpha=10^{-4}$, which
we used for the asteroid belt, results in growth times for pebble accretion of
more than 10 Myr. Hence we set the turbulent stirring to a low value of
$\alpha=10^{-6}$ in the low density model. The high density model has
$\alpha=10^{-5}$. We discuss in the next section how such low values could
arise physically.

We show the resulting size distributions and growth curve of the largest body
at 25 AU in Figure \ref{f:dNdR_R_Rmax_t_25AU}. The size distributions in the
left panel of the figure both display a steep power law beyond 100 km in
radius. Planetesimals up to 300 km in radius accrete significant amounts of
pebbles. Beyond 300 km a run-away growth sets in where the largest body
detaches from the remaining population (right panel of Figure
\ref{f:dNdR_R_Rmax_t_25AU}). This run-away growth is possible because of the
weak gas friction at 25 AU. Hence run-away pebble accretion is not slowed down
by reaching the strong coupling branch but can instead continue to very large
sizes (see Supplementary Material and Figures \ref{f:dMdt_R_25AU} and
\ref{f:eccentricity_pebble_accretion}).

Altogether it is clear that pebble accretion at 25 AU is much slower than at
2.5 AU, and actually requires very low turbulent activity for significant
growth. But planetesimal growth in such wide orbits apparently also has the
potential to result in run-away accretion up to ice giant sizes. The size
distribution of planetesimals from 100 to 300 km in radius is very steep, in
good qualitative agreement with what is observed in the present Kuiper belt
\cite{Fraser+etal2014}. Our results show that 25 AU can be considered the very
limit to where pebble accretion is significant, and then only for weak
turbulence. Beyond this radius planetesimals maintain their birth sizes (and
size distribution), and hence the very steep size distribution of large
planetesimals in the cold component of the classical Kuiper belt may reflect
the actual exponential cut off in the underlying birth size distribution.

In the solar system, Neptune stopped its outwards migration at 30 AU. Therefore
the primordial Kuiper belt can not have extended much beyond that distance, as
otherwise Neptune would have continued to migrate outwards by scattering
planetesimals \cite{Tsiganis+etal2005}. However, the presence of the cold
population, which appears detached from Neptune, indicates that the Kuiper belt
could simply have transitioned to a much lower surface density around 30 AU.
The planetesimals there did not grow beyond their birth sizes, resulting in a
jump in the planetesimal column density between the regions inside 25 AU which
experienced growth by pebble accretion and the regions outside which did not.

\vspace{0.2cm}
\noindent {\bf \large Discussion}
\vspace{0.2cm}

\noindent {\bf Variation of turbulence strength}

An important and relatively unconstrained parameter in our chondrule accretion
model for the formation of asteroids and planetary embryos is the strength of
the turbulence in the protoplanetary disk, as the turbulent diffusion
coefficient sets the scale-height of the particle layer and the mid-plane
density of the chondrules \cite{Johansen+etal2014}. The asteroid formation
region is located in a region of the protoplanetary disk where the degree of
ionisation is believed to be too low to sustain turbulence driven by the
magnetorotational instability. The ionisation degree in the surface layers may
nevertheless be high enough to drive accretion there. The turbulent motion in
these layers will stir the otherwise laminar mid-plane.  Effective turbulent
viscosities between $\alpha=10^{-5}$ and $\alpha=10^{-4}$ have been measured in
the mid-plane in computer simulations of dead zone stirring
\cite{Oishi+etal2007}.

Models of protoplanetary accretion disks that also include the effect of
ambipolar diffusion in the surface layers find that the growth of the
magnetorotational instability is quenched even there, by the lack of coupling
between electrons and neutrals \cite{BaiStone2013}. The magnetic field may
enter a configuration where gas (and angular momentum) is expelled upwards,
while gas connected to the magnetic field lines closer to the mid-plane is
accreted towards the star.  This disk wind accretion leaves the mid-plane
completely laminar. The sedimentation of dust will nevertheless cause a mild
stirring of the mid-plane \cite{Carrera+etal2014}. The stirring caused by
streaming and Kelvin-Helmholtz instabilities in the dense mid-plane layer of
chondrules will be much milder than the stirring from the active layers, with a
typical value of $\alpha=10^{-6}$ for chondrule-sized particles.

\begin{figure}[!t]
  \begin{center}
    \includegraphics[width=0.9\linewidth]{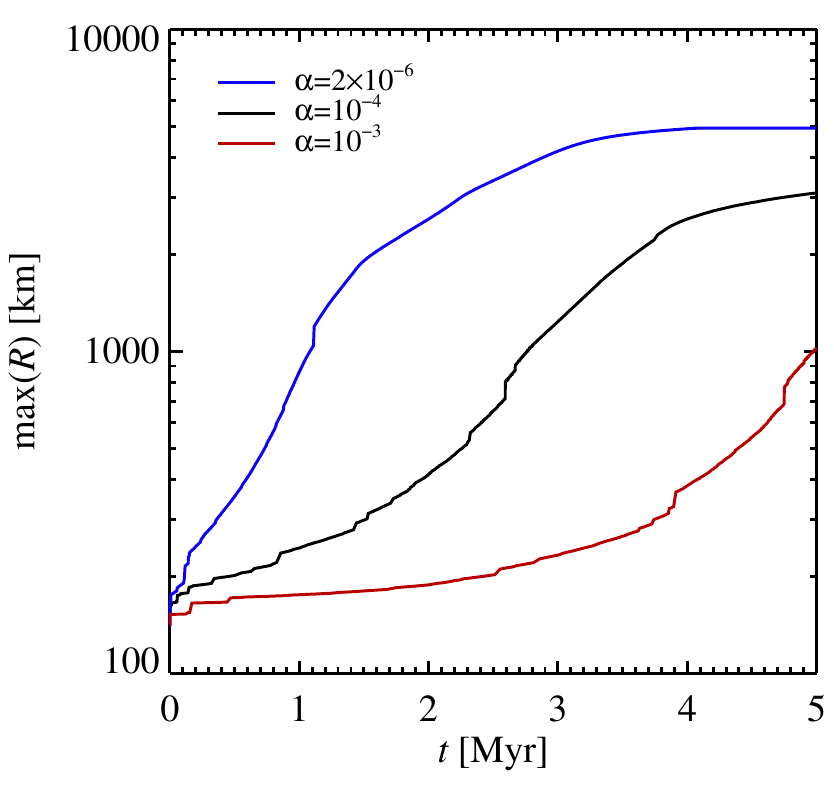}
  \end{center}
  \caption{The maximum planetesimal radius in the asteroid belt versus time,
  for three different values of the turbulent viscosity $\alpha$. Here
  $\alpha=2\times10^{-6}$ represents the strength of turbulence caused by
  streaming instabilities and Kelvin-Helmholtz instabilities in a sedimented
  mid-plane layer of chondrules, $\alpha=10^{-4}$ represents the turbulence
  strength in a dead zone stirred by active surface layers and $\alpha=10^{-3}$
  the turbulence strength caused directly by the magnetorotational instability.
  Turbulent stirring of chondrules sets the scale-height and mid-plane density
  of the chondrule layer and hence dictates the planetesimal growth rate. The
  formation time of the first embryo depends strongly on the degree of
  stirring.}
  \label{f:Rmax_t_deltat}
\end{figure}

The nominal value for the turbulent diffusion coefficient in our model is
$\alpha=10^{-4}$, corresponding to conditions in a dead zone stirred by active
surface layers. To quantify the effect of the turbulent stirring of chondrules
on our results we have run simulations with a lower ($\alpha =2\times10^{-6}$)
and a higher ($\alpha=10^{-3}$) value of the turbulent diffusion coefficient.
The size of the largest planetesimal in the asteroid belt is shown as a
function of time in Figure \ref{f:Rmax_t_deltat}. Clearly, a lower value of the
turbulent diffusion coefficient results in a much higher growth rate of the
planetesimals, as chondrules are allowed to sediment much more strongly to the
mid-plane. For $\alpha=10^{-3}$ the formation of the first embryos is delayed
to after 5 Myr, which is longer than the typical life-times of protoplanetary
disks around young stars.

\noindent {\bf Layered accretion in the asteroid belt}

Our model provides a direct connection between chondrules and asteroid growth.
While early accreted asteroids and their chondrules would melt from the energy
released by the decay of the short-lived $^{26}$Al radionuclide, layers
accreted later than 2 Myr may preserve their pristine, undifferentiated
nature\cite{HeveySanders2006}. Indeed, the existence of differentiated
asteroids overlaid by chondritic crusts has been proposed to explain the
systematic magnetisations of chondrules in the Allende
meteorite\cite{Elkins-Tanton+etal2011,WeissElkins-Tanton2013}, imprinted
from a geodynamo operating in the liquid central region of the parent body.
The asteroid Lutetia has been suggested to be partially differentiated, due to
its high mean density and primitive surface \cite{Patzold+etal2011}.

The characteristic maximum size of planetesimals formed through streaming
instabilities is 50-150 km in radius for column densities comparable to the
Minimum Mass Solar Nebula (Figure \ref{f:size_distribution}). Planetesimals
much smaller than 100 km in radius are relatively inefficient at accreting
chondrules and hence accrete only a thin veneer of chondrules on top of
the primordially formed planetesimal (Figure \ref{f:planetesimal_variation} and
Figure \ref{f:meand_meanphi_R}). Planetesimals forming before 1--2 Myr after
CAIs will likely differentiate, as was the case for the parent bodies of the
differentiated meteorites \cite{Baker+etal2005,Kleine+etal2009}.  Hence we
predict that asteroids currently residing in the knee at the size distribution
at 60 km radii represent primordial, differentiated bodies overlain by a veneer
of chondrules of thickness up to a few dozen kilometers.

Asteroids larger than 100 km in radius, on the other hand, grow mainly by
accretion of chondrules. The initial planetesimal seeds will be covered with
extended layers of narrowly size-sorted chondrules. The Allende meteorite,
whose systematic magnetisation makes the parent body a prime candidate for
layered accretion, has been proposed to originate from the Eos family in the
asteroid belt \cite{Mothe-Diniz+etal2008}. This family is known to display
significant spectral variations between its members
\cite{Mothe-Diniz+etal2008}, indicating partial differentiation and internal
heterogeneity of the 110-km-radius parent body. Another example of an asteroid
family with large internal variation in the spectral properties is the Eunomia
family \cite{WeissElkins-Tanton2013}, believed to originate from a parent body
of 150 km in radius. Intermediate-sized asteroids of 100--200 km in radius
appear to be the best candidates for producing such families with large
internal variation. Bodies of this size range have comparable mass fractions in
the differentiated planetesimal seed and in the accreted chondrule layers.

Many other asteroid families display internal albedo distributions that are
much narrower than the spread of albedos across families
\cite{Mothe-Diniz+etal2005}, in stark contrast to the heterogenous Eos and
Eunomia families discussed above. The inference of the internal structure of
the parent bodies of such apparently homogeneous families is nevertheless
complicated by the identification of interlopers in (and exclusion from) the
asteroid families. A family member whose albedo is different from the family
mean could in fact be mistaken for an interloper. However, the identification
of asteroid families that appear to originate from internally homogeneous
asteroids is not in conflict with our model. Our results show that many
asteroids in the 50--100 km radius range will have experienced little chondrule
accretion and represent primordial asteroid seeds. Additionally, very large
asteroids of radii larger than 300--400 km must have undergone significant
accretion of chondrules between 2 and 3 Myr after the formation of the seed
planetesimal.  Families formed mainly from the chondrule layers of such large
asteroids would also appear relatively homogeneous. Fragments as large as 100
km in radius could consist of purely chondritic material, and families produced
from those fragments would in turn appear entirely homogeneous.

\noindent {\bf Implications for terrestrial planet formation}

The largest planetesimals in our simulations of the asteroid belt and the
terrestrial planet formation region grow to sizes of several thousand
kilometers, forming the embryos whose mutual collisions drive the subsequent
terrestrial planet formation stage. With only 11\% of Earth's mass, Mars may be
one of these remaining embryos. Mars is inferred to have accreted $2\pm1$
Myr after the first condensations in the solar system
\cite{DauphasPourmand2011}, in good agreement with the formation time-scale of
planetary embryos by chondrule accretion in our simulations. This implies that
chondrule accretion is a driving mechanism ensuring the rapid growth of
planetary embryos in the terrestrial planet formation region. Thus the
production of chondrules may be a key process promoting the assembly of rocky
planets.

\vspace{0.2cm}
\noindent {\bf \large Materials and Methods}
\vspace{0.2cm}

\noindent {\bf Planetesimal formation simulations}

The planetesimal formation simulations were performed with the Pencil Code
(which can be freely downloaded, including modifications done for this work, at
http://code.google.com/pencil-code). A new sink particle algorithm developed
for this project is presented in the Supplementary Material.

Our simulations of planetesimal formation through streaming instabilities
are performed in the local shearing box frame. The simulation box orbits with
the local Keplerian frequency $\varOmega$ at an arbitrary radial distance $r$
from the central star. We fix the box size to a cube of side lengths $L=0.2
H$, where $H$ is the gas scale-height. The box size is chosen in order to
capture the relevant wavelengths of the streaming instability
\cite{YangJohansen2014}.  The gas is initialised with a Gaussian density
profile around the mid-plane, in reaction to the component of stellar gravity
directed towards the mid-plane.  The temperature is assumed to be a constant
set through the sound speed $c_{\rm s}$, which is kept fixed during the
simulation.  Particles are given positions according to a Gaussian density
distribution, with a scale-height of 1\% of the gas scale-height. We assume
that particles in the entire column have sedimented into the simulation domain,
giving a mean solids-to-gas ratio of $\langle \rho_{\rm p} \rangle/\rho_0
\approx 0.25$ for the chosen metallicity $Z=0.01$. The particles have uniform
sizes given by the Stokes number ${\rm St}=\varOmega t_{\rm f}=0.3$, where
$\tau_{\rm f}$ is the friction time of the particles, corresponding to
approximately 25-cm-sized particles at 2.5 AU. It has previously been shown
that the consideration of a range of particles sizes gives similar results to
the monodisperse case \cite{Johansen+etal2007}.

The strength of the particle self-gravity in a scale-free local box simulation
is set by the non-dimensional parameter
\begin{equation}
  \varGamma = \frac{4 \pi G \rho_0}{\varOmega^2} \approx 0.036 \left( \frac{r}{2.5\,{\rm AU}} \right)^{1/4} \, .
\end{equation}
Here $G$ is the gravity constant, $\rho_0$ is the gas density in the mid-plane
and $\varOmega$ is the Keplerian frequency at the chosen radial distance from
the star. The scaling with semi-major axis is based on the properties of the
Minimum Mass Solar Nebula \cite{Hayashi1981}. The
parameter $\varGamma$ multiplied by the non-dimensional particle density
$\rho_{\rm p}/\rho_0$ enters the Poisson equation for self-gravity, $\nabla^2
\varPhi = 4 \pi G \rho_{\rm p}$, when lengths are measured in units of the
scale-height and times in units of the inverse Keplerian frequency
$\varOmega^{-1}$. The parameter also converts a given mass density of particles
in a grid cell to an actual mass, since the choice of $\varGamma$ defines the
mid-plane gas density $\rho_0$. Thus both the dynamics and resulting
planetesimal masses are strongly influenced by the value of $\varGamma$ (this
is also evident from the results presented in Figure \ref{f:size_distribution}).

\noindent {\bf Chondrule accretion simulations}

We have developed a statistical code which evolves the masses and orbits of
planetesimals as they accrete chondrules and collide mutually. Details of the
code algorithms, test calculations as well as comparisons to other published
results are presented in the Supplementary Materials and Methods.

\vspace{0.2cm}
\noindent{\bf \large Acknowledgments}
\vspace{0.2cm}

\noindent{\bf General} \\
We are grateful to Glen Stewart, Alessandro Morbidelli, Jeff Cuzzi and
Michiel Lambrechts for stimulating discussions. We would like to thank the
anonymous referees for many helpful suggestions which helped to improve the
paper.

\noindent{\bf Funding} \\
A.J.\ is grateful for the financial support from the European Research Council
(ERC Starting Grant 278675-PEBBLE2PLANET), the Knut and Alice Wallenberg
Foundation and the Swedish Research Council (grant 2010-3710). M.-M.M.L.\ was
partly supported by NASA Origins of Solar Systems grant \#NNX14AJ56G, NSF grant
AST10-09802, and the Alexander von Humboldt-Stiftung. M.B. acknowledges funding
from the Danish National Research Foundation (grant number DNRF97) and by the
European Research Council (ERC Consolidator Grant 616027-STARDUST2ASTEROIDS).
This work was granted access to the HPC resources of Research Centre J\"ulich,
Irish Centre for High-End Computing and Rechenzentrum Garching, made available
within the Distributed European Computing Initiative by the PRACE-2IP (PRACE
project ``PLANETESIM''), receiving funding from the European Community's
Seventh Framework Programme (FP7/2007-2013) under grant agreement no. RI-283493.

\nocite{scilastnote}
\bibliography{bibliography3}
\bibliographystyle{ScienceAdvances.bst}

\clearpage

\onecolumn

\setcounter{figure}{0}
\renewcommand{\thefigure}{S\arabic{figure}}

\section*{Supplementary Materials}

 A.~Johansen, M.-M.~Mac Low, P.~Lacerda, \& M.~Bizzarro

\section{Materials and Methods}

\setcounter{page}{1}

\subsection{Sink particles in the Pencil Code}

\begin{figure*}[!t]
  \begin{center}
    \includegraphics[width=0.8\linewidth]{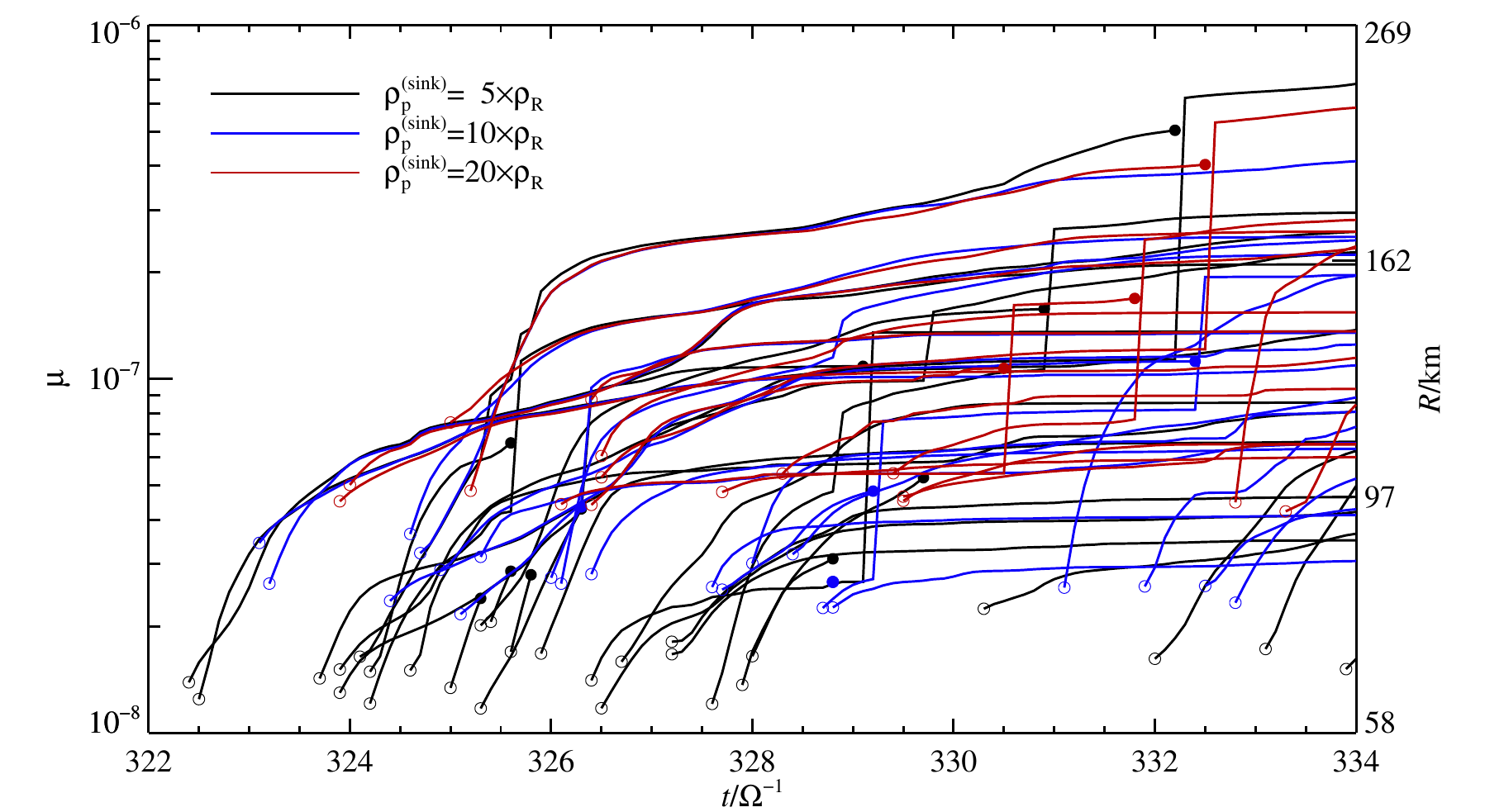}
  \end{center}
  \caption{The non-dimensionless planetesimal
  masses $\mu = \rho_{\rm p} (\delta x)^3$ (here $\rho_{\rm p}$ is the particle
  density represented by the planetesimal in a grid cell and $\delta x$ is the
  size of the cell) and corresponding contracted radii as a function of time
  after self-gravity is turned on. Simulations at $256^3$ grid cells and three
  different values of the sink particle creation threshold are compared.
  Creation of a planetesimal is marked with an empty circle, while the
  destruction is marked with a filled circle. The sizes of the largest
  planetesimals are relatively unaffected by the choice of the threshold. There
  are more mergers of small planetesimals at a low sink particle creation
  threshold, but the planetesimals resulting from those collisions correspond
  well to the massive planetesimals forming at higher creation thresholds.}
  \label{f:mu_t}
\end{figure*}

In order to track the masses of the planetesimals as they form and grow, we
have developed a new sink particle module for the Pencil Code. Above a
threshold particle density value $\rho_{\rm p}=\rho_{\rm p}^{\rm (sink)}$ all
particles in the cell are merged to a single sink particle. This sink particle
does not feel friction with the gas, due to the large size of the
planetesimal.  Any particle coming within the distance $\delta x/2$ of a sink
particle is destroyed ($\delta x$ is the size of the resolution element) and
its mass and momentum added to the sink particle. In Figure \ref{f:mu_t} we
compare the planetesimal masses arising from simulations at $256^3$ with three
different values for the sink particle creation threshold. The figure shows
that the results are relatively unaffected by the choice of sink particle
creation thresholds above 5 times the Roche density. Sink particles allow us to
easily keep track of the temporal evolution and saturation of the planetesimal
masses, and the merging of several million superparticles into a few sink
particles prevents the simulation from slowing down when superparticles cluster
on a subset of the available processors \cite{Johansen+etal2011}.

\subsection{Chondrule accretion simulations}

While classical simulations of planetesimal growth focus on the growth by
gravitationally focused planetesimal-planetesimal
collisions\cite{Bottke+etal2005,Morbidelli+etal2009,Weidenschilling2011}, the
accretion of chondrule-sized objects coupled to the gas via friction can
potentially be much more relevant if planetesimals are born and grow in an
ocean of
chondrules\cite{JohansenLacerda2010,OrmelKlahr2010,LambrechtsJohansen2012}.
``Chondrule accretion'' is an aspect of the more general term ``pebble
accretion'', but chondrules are much smaller than the typically cm-sized
pebbles used in previous studies of pebble accretion.

\subsection{Pebble accretion}

Pebble accretion has two dominant regimes. Planetesimals below a transition
radius of around 1000 km (at 2.5 AU) accrete pebbles embedded in the
sub-Keplerian gas. In the Minimum Mass Solar Nebula\cite{Hayashi1981} the
relative speed between a planetesimal on a circular orbit and the gas is
$\Delta v \approx 50$ m/s.  These pebbles are accreted by the planetesimal from
the Bondi radius of the planetesimal,
\begin{equation}
  R_{\rm B}=G M/(\Delta v)^2 \, ,
\end{equation}  
provided that the friction time of the pebble matches the time to cross the
Bondi radius.  Larger pebbles are simply scattered by the planetesimal, while
smaller pebbles can not be pulled away quickly enough from the almost straight
streamlines of the gas. The finite size of the planetesimal leads to additional
complexities, namely gravitational focusing of loosely coupled pebbles; and
the Stokes flow of the gas around the planetesimal, which affects the
trajectories of strongly coupled particles with impact parameters similar to or
smaller than the radius of the planetesimal.

In order to model all these effects correctly we have solved for the
trajectories of a large number of pebble-planetesimal combinations, for
different values of the friction time $t_{\rm f}$ and the planetesimal radius
$R$. The planetesimal is kept
stationary at $(x,y)=(0,0)$, while the pebble moves under the influence of
planetesimal gravity and gas drag. Pebbles enter the simulation domain at a
large, positive value of $x$ and with impact parameter $b=y_0$. The gas
velocity field is kept fixed and follows the Stokes solution
\cite{Batchelor2000}
\begin{eqnarray}
  u_r &=& -\Delta v \cos(\theta) \left( 1 - \frac{3}{2}\frac{R}{r} + \frac{1}{2}
  \frac{R^3}{r^3} \right) \, , \\
  u_\theta &=& +\Delta v \sin(\theta) \left( 1 - \frac{3}{4}\frac{R}{r} -
  \frac{1}{4} \frac{R^3}{r^3} \right) \, .
\end{eqnarray}
Here $r$ and $\theta$ are the polar coordinates, with $r$ denoting the distance
from the origin and $\theta$ the angle between the $x$-axis and the $y$-axis,
so that the gas flow at large $r$ is uniform along the $x$-direction $\vc{u} =
-\Delta v \hat{\vc{x}}$. Collisions with the planetesimal are treated as
instantaneous, conserving the total momentum as well as the speed component
parallel to the surface and a fraction $c$ of the speed component perpendicular
to the surface, with $c$ denoting the coefficient of restitution. The pebble is
assumed to be accreted after colliding with the planetesimal 5 times or after
orbiting the planetesimal 5 times. The latter criterion is convenient for
planetesimals with large Bondi radii where the pebble enters a slowly decaying
orbit around the planetesimal.

We show the accretion radius $R_{\rm acc}$ (the impact parameter required for
accretion) for (a) perfect sticking and (b) a coefficient of restitution of
$c=0.5$ in Figure \ref{f:bacc_tauf}. The accretion radius depends on both the
friction time (normalised by the Bondi time $t_{\rm B}=R_{\rm B}/\Delta v$ in
Figure \ref{f:bacc_tauf}) and the planetesimal size (normalised by the Bondi
radius). We discuss the general trends in the accretion radius curve here:
\begin{figure*}[!t]
  \begin{center}
    \includegraphics[width=0.45\linewidth]{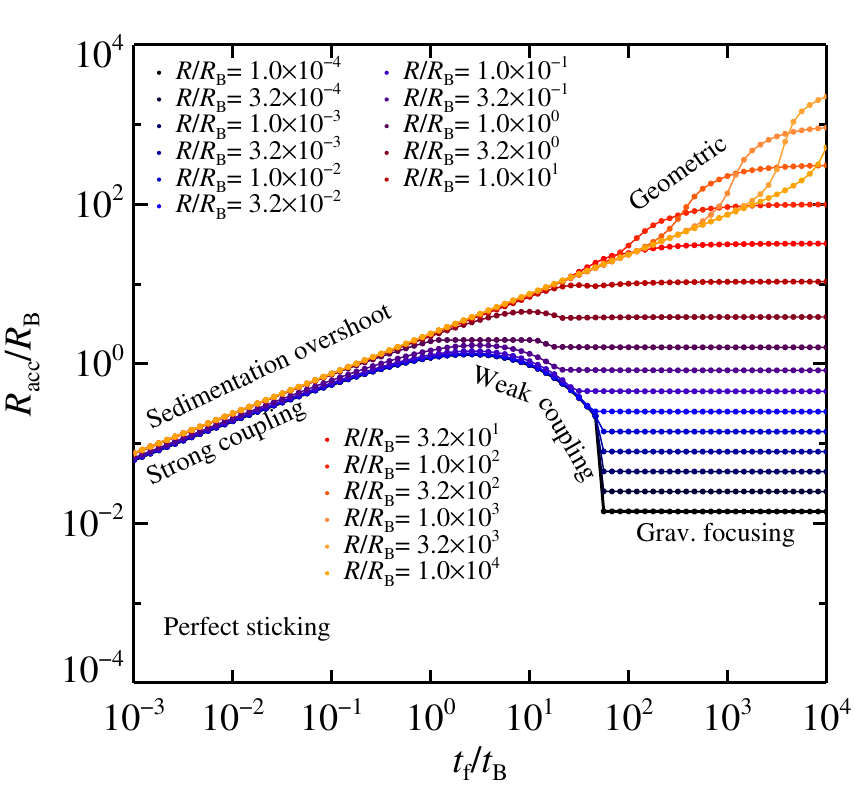}
    \includegraphics[type=pdf,ext=.pdf,read=.pdf,width=0.45\linewidth]{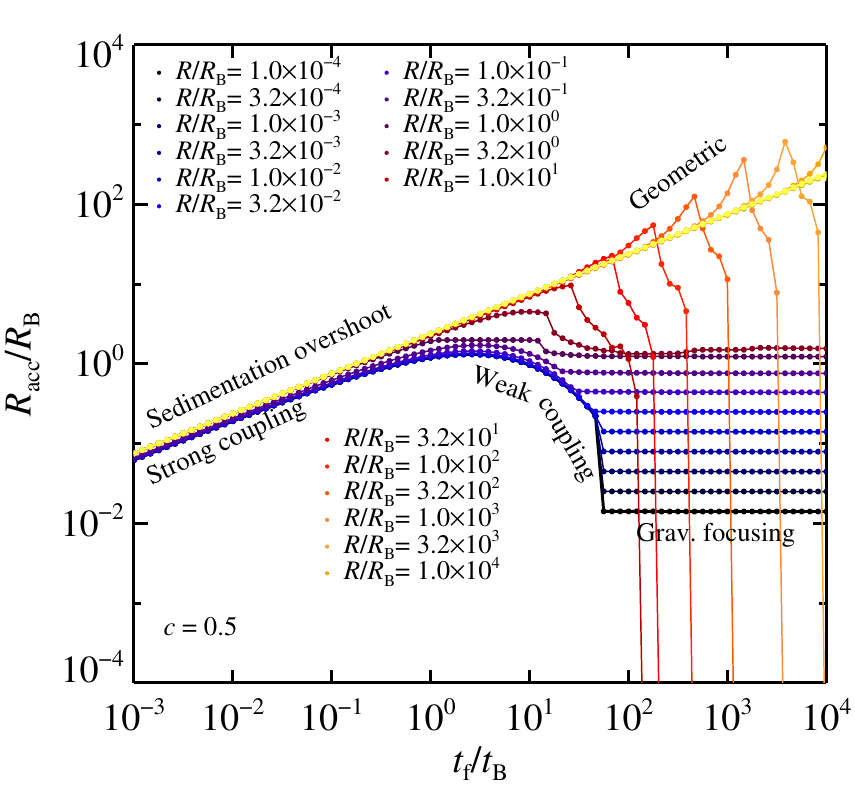}
  \end{center}
  \caption{The accretion radius $R_{\rm acc}$ (i.e.\, the impact parameter
  required for accretion, given in units of the Bondi radius $R_{\rm B}$)
  versus the particle friction time $t_{\rm f}$ (in units of the Bondi
  time-scale $t_{\rm B}$), for planetesimal radii $R$ between $10^{-4} R_{\rm
  B}$ and $10^4 R_{\rm B}$.  The left panel shows the results when assuming
  perfect sticking between pebble and planetesimal. Planetesimals with
  $R<R_{\rm B}$ have three rather distinct branches: strong coupling for short
  friction times, weak coupling for longer friction times and finally
  gravitational focusing for the longest friction times. Planetesimals with
  $R>R_{\rm B}$ radius have two branches: sedimentation overshoot for short
  friction times (which follows the strong coupling scaling but is physically
  distinct) and geometric accretion for friction times longer than the time to
  pass the planetesimal. The consideration of the outcome of the
  pebble-planetesimal collision, with a coefficient of restitution $c=0.5$, is
  shown the right panel. This strongly limits geometric accretion onto small
  planetesimals.}
  \label{f:bacc_tauf}
\end{figure*}

{\it Large Bondi radius.} The gas flow around the planetesimal and the
coefficient of restitution play no role in determining the accretion radius
when the planetesimal is much smaller than the Bondi radius. This is the case
for large planetesimals and/or low relative pebble-planetesimal speeds. The
three sections on the growth curve in Fig.\ \ref{f:bacc_tauf} are: (i) Strong
coupling when the pebble couples to the gas on a much shorter time-scale than
the Bondi time-scale.  Here the accretion radius is proportional to $R_{\rm B}
\sqrt{t_{\rm f}/t_{\rm B}}$. (ii) Weak coupling when the pebble couples to the
gas on much longer time-scales than the Bondi time-scale. Here the accretion
radius drops rapidly with increasing friction time, with no simple analytical
expression.  (iii) Gravitational focussing for the largest particles when the
weak coupling branch drops below the gravitational focussing radius.

{\it Small Bondi radius.} A small planetesimal has Bondi radius much smaller
than the planetesimal radius. This limit has two distinct branches: (i)
Sedimentation overshoot for strongly coupled pebbles. These pebbles follow the
Stokes flow around the planetesimal, but manage to sediment on to the surface
for very low impact parameters where the gas flow comes very close to the
planetesimal surface. (ii) Geometric accretion for weakly coupled pebbles that
do not react to the modified gas flow close to the planetesimals. The latter
branch in turn is strongly affected by the coefficient of restitution. Perfect
sticking leads to accretion onto the entire projected surface, while
consideration of the collision outcome requires very low impact parameters to
remain bound after the collision.

Below we discuss some important aspects of the sedimentation overshoot and
geometric branches.

{\it Sedimentation overshoot.} Small particles are carried with the gas around
the planetesimals. The frictional acceleration between gas and particles is
nevertheless not instantaneous but operates on the time-scale $t_{\rm f}$.
This can lead to accretion of particles that sediment onto the planetesimal
surface during the transport around the planetesimal. The gas streamlines at
a low impact parameter of $y_0=b\ll R$ move around the planetesimal at the
distance $\delta r \sim b$, at the approximate speed
\begin{equation}
  u_\theta \sim \Delta v \frac{b}{R} \, .
\end{equation}
Ignoring factors of order unity, the time to move around the planetesimal is
then
\begin{equation}
  \tau_\theta \sim \frac{R}{u_\theta} \sim \frac{R^2}{b \Delta v} \, .
\end{equation}
During this time the pebble sediments towards the planetesimal at the terminal
velocity
\begin{equation}
  v_{\rm t} = t_{\rm f} \frac{G M}{R^2} \, .
\end{equation}
The criterion for reaching the surface is $v_{\rm t} \tau_\theta \sim b$.
This in turn yields
\begin{equation}
  b^2 \sim t_{\rm f} \frac{G M}{\Delta v} \, .
\end{equation}
Dividing by the squared Bondi radius finally yields the criterion for accretion
\begin{equation}
  \frac{b^2}{R_{\rm B}^2} \sim \frac{t_{\rm f}}{t_{\rm B}} \, .
\end{equation}
This expression is equivalent to the strong coupling limit of pebble
accretion\cite{LambrechtsJohansen2012}, although the physics of the accretion
via sedimentation overshoot is completely different.

{\it Geometric accretion.} The requirement that the pebble remains bound after
the collision can be found geometrically as
\begin{equation}
  R_{\rm acc} = R \sqrt{\frac{(v_{\rm e}/\Delta v)^2 - c^2}{1-c^2}} \, .
\end{equation}
Pebbles entering with an impact parameter less than $R_{\rm acc}$ are thus
accreted, while pebbles with a larger impact parameter collide once with the
planetesimal and then fly off to infinity. The accretion radius is only
positive for $c < v_{\rm e}/\Delta v = \sqrt{2} R_{\rm B}/R$. Hence
planetesimals with Bondi radii smaller than their size can not accrete pebbles
which couple to the gas on time-scales longer than the time-scale to pass the
planetesimal (unless $c$ is extremely low).

\subsection{Eccentricity and inclination}

The growing asteroids experience mutual gravitational encounters that excite
the eccentricities and inclinations of the population. This affects the pebble
accretion rate since the relative speed between a planetesimal and the
sub-Keplerian pebble flow changes with position in the orbit for non-zero
eccentricities and inclinations.

\begin{figure}[!t]
  \begin{center}
    \includegraphics[width=0.45\linewidth]{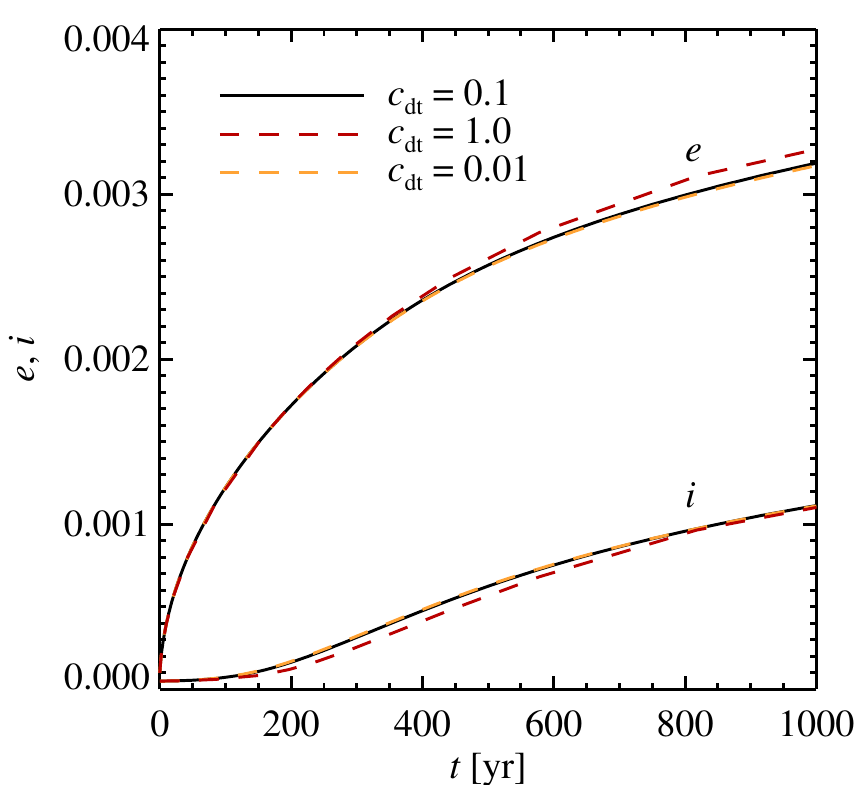}
  \end{center}
  \caption{The evolution of eccentricity $e$ and inclination $i$ of 1000
  planetesimals with mass $M=10^{24}$ g located at 1 AU with a surface density
  of 10 g/cm$^2$, for three values of the time-step. Here $c_{\rm dt}$
  denotes the time-step relative to the collision time-scale. Stirring and
  dynamical friction are implemented using the scheme described in Ohtsuki et
  al.\ (2002). The results are qualitatively similar to Figure 4 of that
  paper, although the $e$ and $i$ evolve a bit slower in our simulation. We
  attribute this to a difference in the set up of the simulations: while we
  initially set $e=10^{-4}$ and $i=5\times10^{-4}$ for all planetesimals,
  Ohtsuki et al.\ (2002) give their planetesimals a Rayleigh distribution
  around those values. We use $c_{\rm dt}=0.1$ in the actual planetesimal
  growth simulations.}
  \label{f:OSI02_test}
\end{figure}

\begin{figure}[!t]
  \begin{center}
    \includegraphics[width=0.45\linewidth]{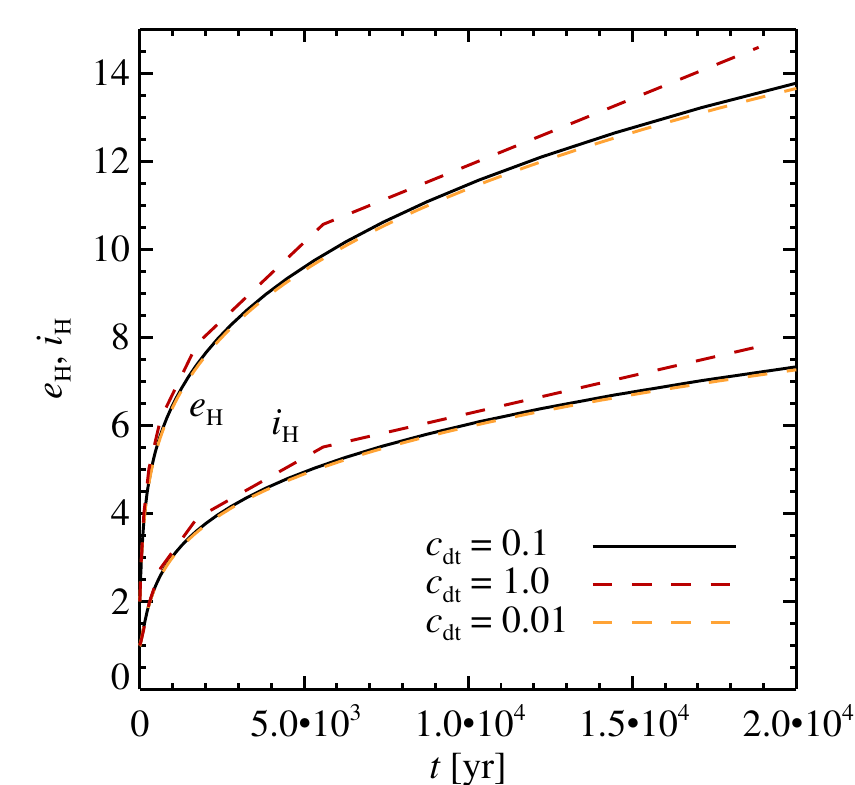}
  \end{center}
  \caption{The evolution of eccentricity $e$ and inclination $i$ of 800
  planetesimals with mass $M=10^{24}$ g located at 1 AU with a surface density
  of 10 g/cm$^2$. The eccentricity and the inclination are normalised by their
  Hill values $[2 M_{\rm p}/(3 M_\star)]^{1/3}$. Results are qualitatively
  similar to Figure 2 of Stewart \& Ida (2000). As in Figure
  \ref{f:OSI02_test}, a time-step parameter of $c_{\rm dt}=0.1$ provides a good
  compromise between precision and speed; hence we use $c_{\rm dt}=0.1$ for
  planetesimal stirring in the planetesimal growth simulations.}
  \label{f:SI00_test}
\end{figure}

The temporal evolution of the eccentricity $e$ and the inclination $i$ is
calculated using the analytical approximations of Ohtsuki et al.\
\cite{Ohtsuki+etal2002}, constructed to match the results of $N$-body
simulations for both low and high eccentricities. The analytical evolution
equations for $e$ and $i$ give the excitation as well as dynamical friction of
a binned population of planetesimals. We show in Figure \ref{f:OSI02_test}
the result of a test problem defined in Ohtsuki et al. Here we
consider 1000 equal-mass planetesimals with initial eccentricity of $10^{-4}$
and inclination of $5\times10^{-5}$. The evolution of $e$ and $i$ follows the
general curve in Figure 4 of Ohtsuki et al., but the evolution is a
bit slower. We attribute this to a difference in the tests. In Ohtsuki et al.\
the planetesimals are given a Rayleigh distribution of $e$ and $i$
initially, while we assign a constant value equal to the mean of that Rayleigh
distribution. A similar test problem with 800 planetesimals is shown in Figure
\ref{f:SI00_test}, with good agreement with Figure 2 of Stewart \& Ida
\cite{StewartIda2000}.

In principle we could use all the planetesimal particles in the code as natural
bins and evolve each planetesimal with the dynamical excitation contribution
from all other planetesimals. However, this would take unfeasibly long time to
compute when there are millions of planetesimal particles. Instead we bin the
planetesimals by their radius and consider the dynamical evolution of the
smallest and largest planetesimal in each bin. Gravitational stirring is only
considered between these anchor planetesimals. The resulting values of ${\rm
d}e^2/{\rm d}t$ and ${\rm d}i^2/{\rm d}t$ are then multiplied by the total
number of planetesimals in the stirring bin, divided by the number of
planetesimals that were actually considered (the two anchors). The evolution
of $e$ and $i$ is finally interpolated from the anchor planetesimals to all the
other planetesimals at the end of the time-step.

The eccentricity and inclination of the orbit can now be used to calculate the
pebble accretion rate over a full orbit of the planetesimal. For small
eccentricities and inclinations the orbit can be considered in the local
coordinate frame corotating with the Keplerian flow at a given distance $r_0$
from the star. The coordinate system is oriented such that the $x$-axis points
radially away from the star, the $y$-axis along the orbital direction of the
gas and the $z$-axis perpendicular to the plane of the disk, along the rotation
vector of the disk. The epicyclic motion of a planetesimal with eccentricity
$e$ in such a frame is
\begin{eqnarray}
  v_x(t) &=& v_{\rm e} \cos(\varOmega t) \, , \label{eq:vxp} \\
  v_y(t) &=& - \frac{1}{2} v_{\rm e} \sin(\varOmega t) \, , \label{eq:vyp} \\
  v_z(t) &=& v_{\rm i} \cos(\varOmega t) \label{eq:vzp} \, .
\end{eqnarray}
Here $v_{\rm e} = e v_{\rm K,0}$ and $v_{\rm i} = i v_{\rm K,0}$ are the
eccentricity and inclination speeds of the orbit, respectively, measured
relative to the Keplerian speed at the centre of the frame, and $(v_x,v_y)$ is
the velocity is measured relative to the {\it local} Keplerian motion at the
instantaneous position of the planetesimal. Relative to $v_{\rm K,0}$ the
azimuthal velocity is
\begin{equation}
  \tilde{v}_y(t) = - 2 v_{\rm e} \sin(\varOmega t) \, .
\end{equation}
The velocity of the incoming chondrules is $v_y' = -\Delta v$ relative to the
local Keplerian velocity. Hence the relative speed between planetesimal and
chondrules is
\begin{equation}
  v_{\rm rel} = \sqrt{ [v_{\rm e} \cos(\varOmega t)]^2 + [ -(1/2) v_{\rm e}
  \sin(\varOmega t) + \Delta v]^2 + [v_{\rm i} \cos(\varOmega t)]^2} \, .
\end{equation}
The relative speed is lowest at aphelion ($\varOmega t = \pi/2$) and highest at
perihelion ($\varOmega t = 3 \pi/2$). Since the Bondi radius scales as $R_{\rm
B} \propto 1/v_{\rm rel}^2$, the eccentric motion strongly affects the
accretion rate. On the strong coupling branch of pebble accretion one can
nevertheless show that the eccentric orbit will not affect the mass accretion
rate, because in fact the accretion rate is independent of $v_{\rm rel}$,
\begin{equation}
  \dot{M}_{\rm SC} \propto \frac{t_{\rm f}}{t_{\rm B}} R_{\rm B}^2 v_{\rm
  rel} \propto \frac{t_{\rm f}}{G M / v_{\rm rel}^3} \frac{G^2 M^2}{v_{\rm
  rel}^4} v_{\rm rel} \propto G M t_{\rm f} \, .
\end{equation}
At aphelion the relative speed between chondrule and planetesimal can even
approach zero for $(1/2) v_{\rm e} \sim \Delta v$. At this point the Bondi
radius is no longer the relevant accretion radius as the Hill speed of the
planetesimal is higher than the relative chondrule-planetesimal speed, and the
planetesimal thus enters short periods of Hill accretion at aphelion (our
implementation of Hill accretion is explained in section \ref{s:hill}).

\begin{figure}
  \begin{center}
    \includegraphics[width=0.45\linewidth]{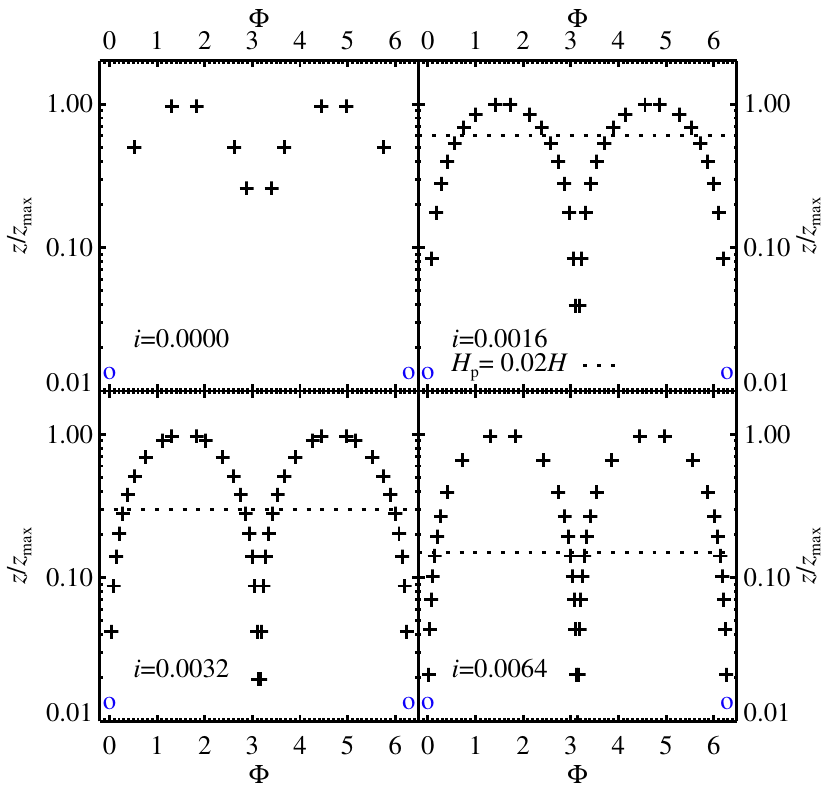}
  \end{center}
  \caption{Phase points covering planetesimal
  orbits with inclinations $i=0$, $i=0.0016$, $i=0.0032$ and $i=0.0064$. We
  construct phase arrays to cover both the full planetesimal orbit and the
  scale-height of the chondrule layer (here $H_{\rm p}=0.02 H$). The height in
  the planetesimal orbit is normalised by the maximum height for the given
  inclination. The dotted line indicates the scale-height of the chondrule
  mid-plane layer, clearly resolved for all choices of the inclination.}
  \label{f:phases_pebble_accretion}
\end{figure}

We take into account the eccentric and inclined orbit of the planetesimal by
sampling the accretion rate at a number of phases in the orbit. The choice of
phase points is a balance between the need to sample both the scale height
$H_{\rm p}$ of the chondrule layer, where chondrule densities are highest, as
well as the planetesimal orbit over which the relative planetesimal-chondrule
speed, and hence the accretion radius, varies. We set the first phase point at
$\Phi=\varOmega t=0$. The distance to the next phase point is taken as the
minimum of the two functions
\begin{eqnarray}
  \Delta \Phi_1 &=& \frac{\exp(z/H_{\rm c}) H_{\rm c}}{z_{\rm orb}} \frac{2
  \pi}{N_1} \, , \\
  \Delta \Phi_2 &=& \frac{2 \pi}{N_2} \, .
\end{eqnarray}
Here $z_{\rm orb}$ is the maximal height of the planetesimal relative to the
mid-plane. This gives phases that cover the full planetesimal orbit, with
additional resolution elements added within the chondrule scale height.  This
way we avoid having to use a very large number of points for highly inclined
orbits.  We show examples of phase arrays in Figure
\ref{f:phases_pebble_accretion} for $N_1=48$ and $N_2=12$.

\begin{figure}[!t]
  \begin{center}
    \includegraphics[type=pdf,ext=.pdf,read=.pdf,width=0.7\linewidth]{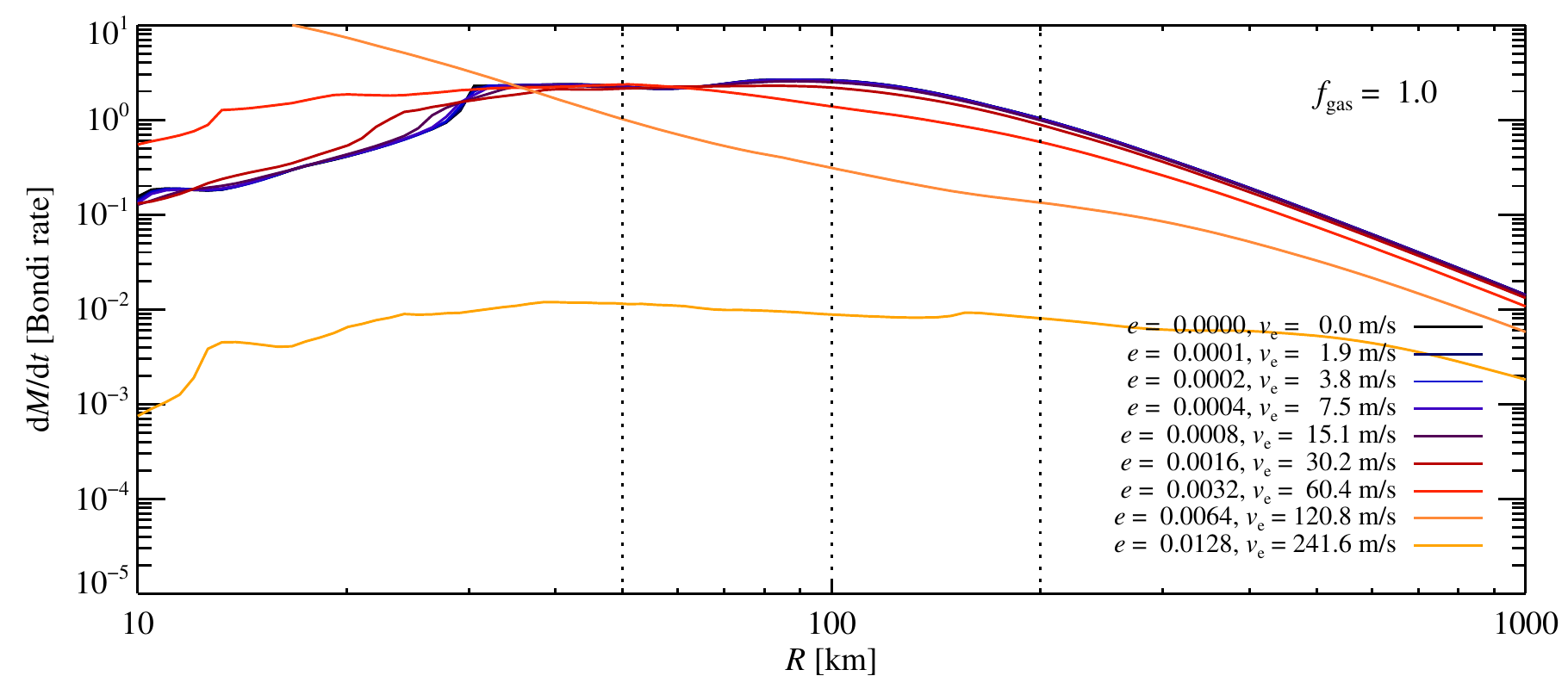}
    \includegraphics[type=pdf,ext=.pdf,read=.pdf,width=0.7\linewidth]{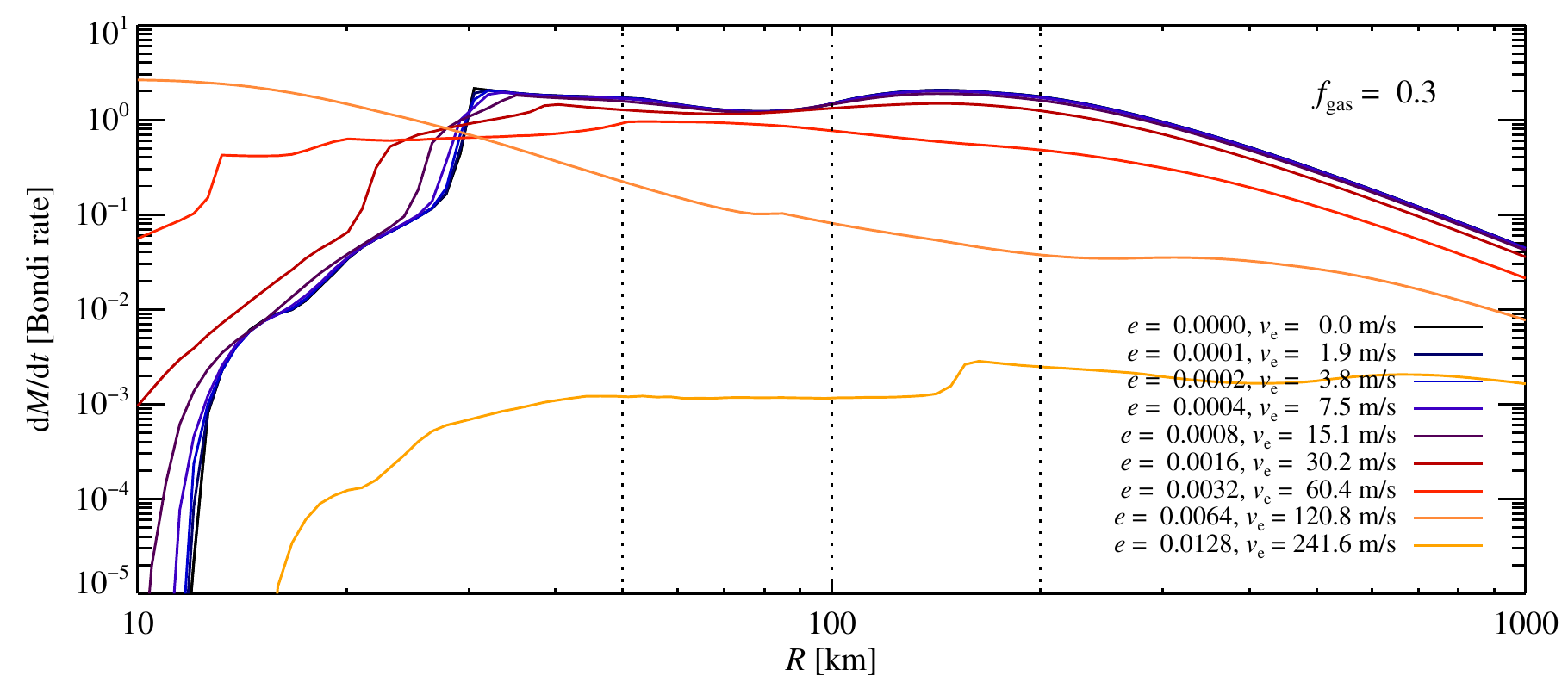}
    \includegraphics[type=pdf,ext=.pdf,read=.pdf,width=0.7\linewidth]{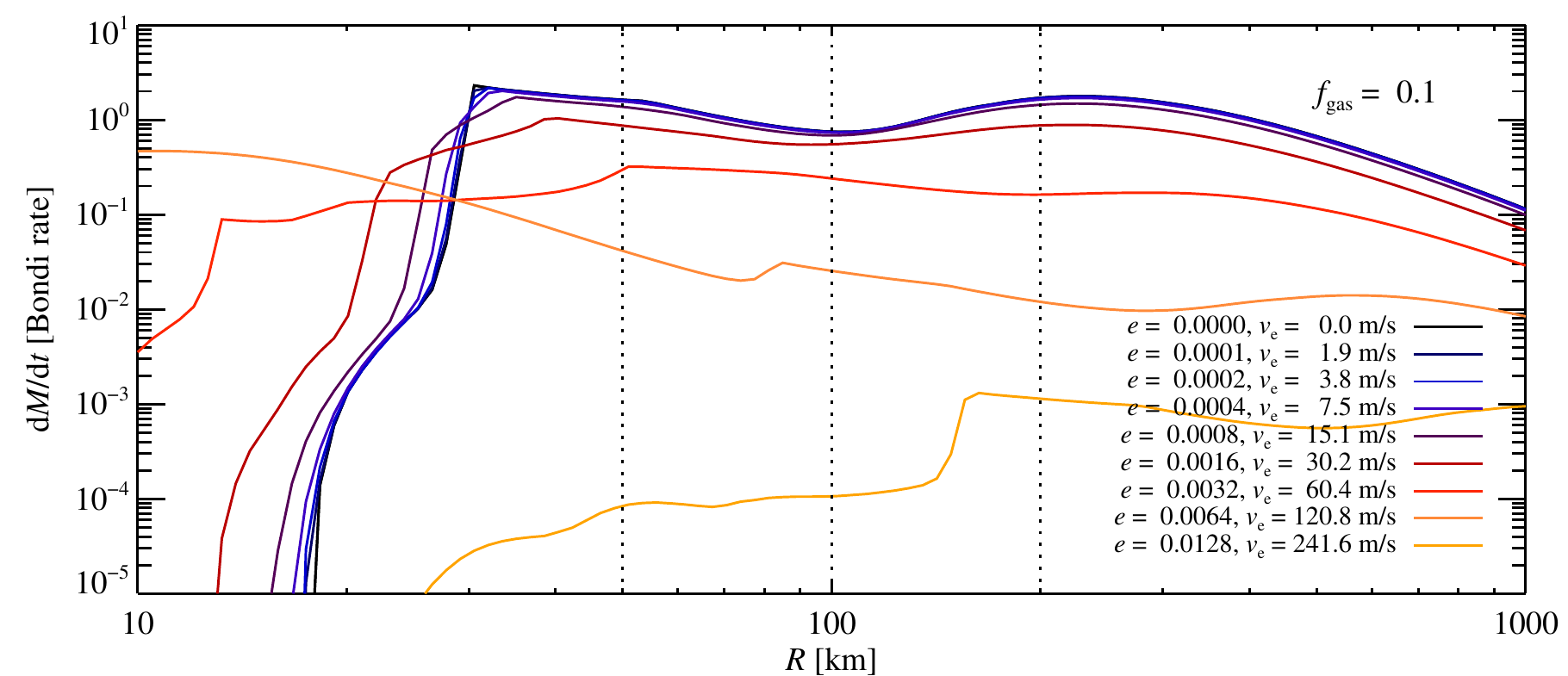}
  \end{center}
  \caption{The effect of eccentricity and inclination on the chondrule
  accretion rate, at an orbital distance of 2.5 AU. The $x$-axis shows the
  planetesimal sizes and the $y$-axis the pebble accretion rate normalised to
  the accretion rate from the full Bondi radius. Dotted lines mark 50 km, 100
  km and 200 km. We use here the approximation $i=e/2$. Increasing the orbital
  eccentricity can actually increase the accretion rate on bodies smaller than
  100 km in radius, as the orbital speed in aphelion matches the sub-Keplerian
  speed of the chondrules which leads to very high accretion rates. Larger
  planetesimals are affected negatively beyond the threshold $(e/2) v_{\rm K} =
  \Delta v$ as the large particles necessary for efficient accretion at
  aphelion are not present in the disk.  The top plot shows results for the
  column density of the Minimum Mass Solar Nebula at 2.5 AU, the middle plot
  $0.3$ times the MMSN and the bottom plot $0.1$ times the MMSN.}
  \label{f:eccentricity_pebble_accretion}
\end{figure}

In Figure \ref{f:eccentricity_pebble_accretion} we show the mass accretion rate
of planetesimals at 2.5 AU as a function of the eccentricity of the orbit. We
use the approximation that $i=e/2$ and measure the accretion rate relative to
that at $e=i=0$.  Increasing the eccentricity actually increases the geometric
accretion rate on small planetesimals (below 50 km in radius), since the
relative planetesimal-chondrule speed increases and hence the mass flux
increases proportionally to the speed.  Larger planetesimals are affected
negatively by eccentricity. These planetesimals accrete small chondrules at
very high rate on circular orbits. The accretion rate at the aphelion of an
eccentric orbit does not benefit strongly from the decreased relative
planetesimal-chondrule speed at that phase, since the larger Bondi radius
requires very large chondrules, which are not present in the disk.  Instead the
planetesimal enters the strong coupling branch at aphelion, while the accretion
rate is strongly reduced at perihelion.  The overall result is a reduction in
the accretion rate. Even larger planetesimals are affected less and less by the
increase in eccentricity. These large planetesimals already accrete on the
strong coupling branch, which is relatively unaffected by the relative speed
and hence by the eccentricity.
\begin{figure}[!t]
  \begin{center}
    \includegraphics[width=0.7\linewidth]{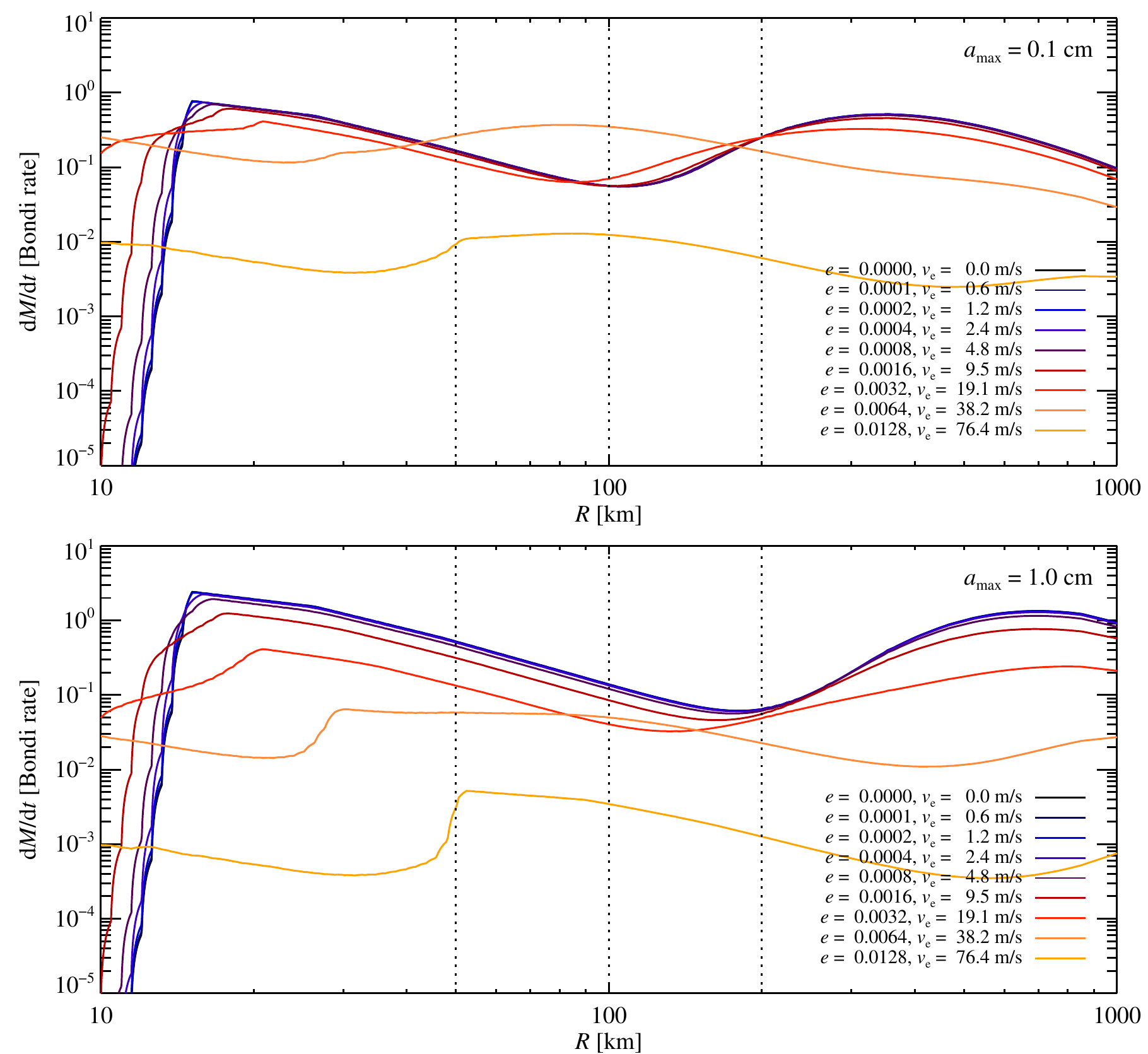}
  \end{center}
  \caption{The accretion rate of planetesimals at 25 AU, as a function of
  the planetesimal size. The accretion rate is normalised by the Bondi rate.
  The top panel shows the accretion rate for pebbles up to 1 mm in radius,
  while the bottom panels shows the accretion rate for pebbles up to 1 cm in
  radius.  The accretion rate beyond 200 km in radius is flat or increasing
  here in the outer regions of the protoplanetary disk. This is in contrast to
  the situation in the asteroid belt where the lack of chondrules of larger
  than mm size slows down the run-away accretion of large planetesimals (Figure
  \ref{f:eccentricity_pebble_accretion}). The presence of mm-cm pebbles in the
  outer protoplanetary disk can drive a run-away accretion where the most
  massive planetesimals detach from the rest of the population (see Figure
  \ref{f:dNdR_R_Rmax_t_25AU}).}
  \label{f:dMdt_R_25AU}
\end{figure}

Figure \ref{f:dMdt_R_25AU} illustrates the mass accretion rates of
planetesimals at 25 AU for different values of their eccentricity. The
normalised accretion rates in Figure \ref{f:dMdt_R_25AU} are flat or increasing
beyond 200 km in radius, indicating a very steep increase in the accretion rate
with size. This is in contrast to the situation in the asteroid belt where the
run-away accretion of chondrules is stopped by the lack of chondrules larger
than mm in size (Figure \ref{f:eccentricity_pebble_accretion}).

\subsection{Stratification integral}

\begin{figure}[!t]
  \begin{center}
    \includegraphics[width=0.45\linewidth]{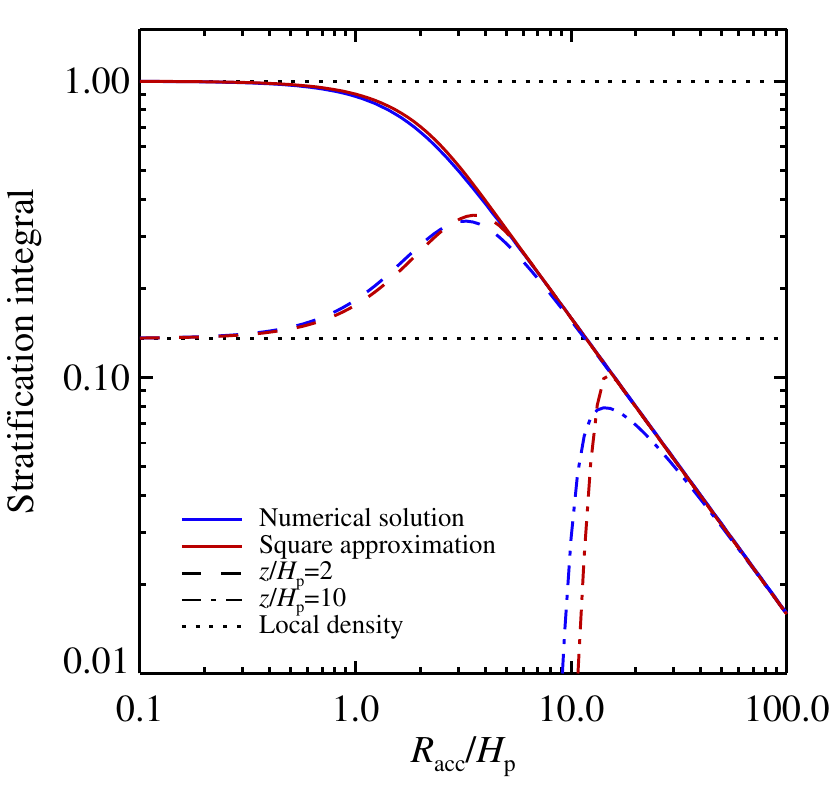}
  \end{center}
  \caption{The stratification integral (i.e., the mean particle density over
  the accretion cross section) as a function of the accretion radius of the
  planetesimal. Numerical integration of the Gaussian stratification over a
  circle (blue lines) is expensive, so we use instead a square approximation to
  the integral (red lines). Full lines indicate a planetesimal in the
  mid-plane, dashed lines a planetesimal at two times the chondrule
  scale-height above the mid-plane, and dash-dotted lines a planetesimal at ten
  times the chondrule scale-height. The square approximation is very precise
  both when the accretion radius is much lower or much higher than the
  chondrule scale-height.}
  \label{f:stratification_integral}
\end{figure}

Calculation of the mass accretion rate requires knowledge of the accretion
radius as well as the chondrule density averaged over the accretion radius. The
stratification integral $S$ is defined as the mean chondrule density normalised
by the chondrule density in the mid-plane. The stratification integral for a
planetesimal with accretion radius $R_{\rm acc}$ located at the height $z_0$
over the mid-plane is
\begin{equation}
  S = \frac{1}{\pi R_{\rm acc}^2} \int_{z_0-R_{\rm acc}}^{z_0+R_{\rm acc}}
  \exp[-z^2/(2 H_{\rm p}^2)] 2 \sqrt{R_{\rm acc}^2-(z-z_0)^2} {\rm d} z \, .
  \label{eq:s}
\end{equation}
This expression is obtained by summing over lines of constant $z$ and hence
constant chondrule density. There is no simple analytical solution to equation
(\ref{eq:s}). Tabularisation of the numerical solution requires interpolation
in both $R_{\rm acc}$, $z_0$ and the angle of incidence of the chondrule flow
$\theta$ (see below). We therefore an approximation that allows the integral to
be calculated analytically. This {\it square} approximation integrates the
chondrule density over a square instead of a circle. This yields the solvable
integral
\begin{equation}
  S_{\rm square} = \frac{1}{4 R_{\rm acc}^2} \int_{z_0-R_{\rm acc}}^{z_0+R_{\rm
  acc}} \exp[-z^2/(2 H_{\rm p}^2)] 2 R_{\rm acc} {\rm d} z \, .
\end{equation}
The analytical solution is
\begin{equation}
  S_{\rm square} = \frac{H_{\rm p}}{\sqrt{2} R_{\rm acc}}\frac{\sqrt{\pi}}{2}
  \left[ {\rm erf}\left(\frac{z_0 + R_{\rm acc}}{\sqrt{2} H_{\rm p}}\right) -
  {\rm erf}\left(\frac{z_0 - R_{\rm acc}}{\sqrt{2} H_{\rm p}}\right) \right]
  \, .
  \label{eq:ssquare}
\end{equation}
The square approximation tends to underestimate the chondrule density because
of the inclusion of low-density corners in the square. Therefore we decrease
the size of the square by replacing $R_{\rm acc}$ in equation
(\ref{eq:ssquare}) with $R_{\rm acc}'=f R_{\rm acc}$. We have found $f=0.79$ to
give a much better fit to the numerical integral than $f=1$. In 
Figure \ref{f:stratification_integral} we show how the square
approximation compares to the full numerical integration of the accretion
radius over the Gaussian stratification.

Planetesimals at very high inclinations encounter the chondrule flow at an
angle $\theta$ which is $\neq 0 ^\circ$, measured relative to the vertical. The
angle of incidence is found through
\begin{equation}
  \theta = {\rm acos}\left(\frac{v_z^{\rm (rel)}}{v_{\rm rel}}\right)
\end{equation}
Here $v_{\rm rel}=|\vc{v}_{\rm rel}|=|\vc{v}_{\rm p}-\vc{v}_{\rm c}|$ is the
vectorial relative speed between the planetesimal (with velocity $\vc{v}_{\rm
p}$) and the chondrules (with velocity $\vc{v}_{\rm c}=-\Delta v
\hat{\vc{y}}$). The components of $\vc{v}_{\rm p}$ are given in equations
(\ref{eq:vxp})-(\ref{eq:vzp}). The consideration of the angle of incidence
changes the stratification integral to
\begin{equation}
  S = \frac{1}{\pi R_{\rm acc}^2} \int_{z_0-R_{\rm acc} \cos
  \theta}^{z_0+R_{\rm acc} \cos \theta}
  \exp[-z^2/(2 H_{\rm p}^2)] 2 \sqrt{R_{\rm acc}^2-\left(\frac{z-z_0}{\cos
  \theta}\right)^2} \frac{{\rm d} z}{\cos \theta} \, .
\end{equation}
The square approximation changes to
\begin{equation}
  S_{\rm square} = \frac{H_{\rm p}}{\sqrt{2} R_{\rm acc}}
  \frac{\sqrt{\pi}}{2} \frac{1}{\cos \theta} \left[ {\rm erf}\left(\frac{z_0 +
  R_{\rm acc} \cos \theta}{\sqrt{2} H_{\rm p}}\right) - {\rm
  erf}\left(\frac{z_0 - R_{\rm acc} \cos \theta}{\sqrt{2} H_{\rm p}}\right)
  \right] \, .
\end{equation}
The results presented in this paper are obtained with the square approximation
including the correction for the encounter angle.

\subsection{Transition to Hill accretion}
\label{s:hill}

Bondi accretion is only valid for planetesimals below a transition mass of
approximately 0.1\% of an Earth mass\cite{LambrechtsJohansen2012}. Beyond this
the Hill radius of the growing planetesimal, $R_{\rm H}$, is so large that the
relative speed between chondrule and planetesimal is set by the Hill speed
$v_{\rm H} = \varOmega R_{\rm H}$ rather than the sub-Keplerian speed $\Delta
v$.

We obtain a smooth transition from Bondi accretion to Hill accretion by solving
for the combined speed $v(R_{\rm acc}) = \Delta v + \varOmega R_{\rm acc}$. The
accretion radius $R_{\rm acc}$ depends on the approach speed, so we solve the
equation iteratively from the starting point $v = \Delta v$. This allows us to
also apply the tabulated accretion radius for Bondi accretion (Figure
\ref{f:bacc_tauf}) to Hill accretion, by modifying the approach speed to take
into account the Keplerian shear. We have checked that the resulting accretion
rates on the Hill branch are very similar to the results from hydrodynamical
simulations\cite{LambrechtsJohansen2012}.

\subsection{Eccentricity damping by gas drag and chondrule accretion}

The planetesimals in our model are so large that the quadratic drag force
regime applies, with drag force proportional to the relative speed squared and
with friction time
\begin{equation}
  t_{\rm f} = \frac{6 R \rho_\bullet}{\delta v \rho_{\rm g}} \, .
  \label{eq:tauf}
\end{equation}
The drag force can be applied directly to the eccentricity and inclination,
following Wetherill \& Stewart (1989)\cite{WetherillStewart1989} and Morbidelli
et al.\ (2009)\cite{Morbidelli+etal2009},
\begin{eqnarray}
  \frac{{\rm d}e^2}{{\rm d}t} &=& -\frac{16}{5} \frac{v}{v_{\rm K}^2} \frac{0.5
  \pi \rho_{\rm g} v_{\rm g}^2 R^2}{2 M ( 1 + 0.8 \beta^2)} \, ,
  \label{eq:decc2dtdrag} \\
  \frac{{\rm d}i^2}{{\rm d}t} &=& -3.2 \beta^2 \frac{v}{v_{\rm K}^2} \frac{0.5
  \pi \rho_{\rm g} v_{\rm g}^2 R^2}{2 M ( 1 + 0.8 \beta^2)} \, .
  \label{eq:dinc2dtdrag} 
\end{eqnarray}
Here we use the notation of Morbidelli et al.\ (2009) who define $\beta=i/e$,
$v=v_{\rm K}\sqrt{(5/8)e^2+(1/2)i^2}$ and $v_{\rm g} = \sqrt{v(v+\Delta v)}$.
This formulation damps the eccentricity and inclination to zero on the relevant
friction time-scale, with the term $v+\Delta v$ in $v_{\rm g}$ corresponding to
an orbitally averaged relative speed which would enter equation
(\ref{eq:tauf}). We have checked the analytical evolution of $e$ and $i$
against an $N$-body integration of an eccentric orbit circularised by gas drag
and found excellent agreement.

Chondrule accretion and scattering are two additional damping mechanisms. The
friction time-scale for chondrule accretion with accretion radius $R_{\rm acc}$
is
\begin{equation}
  t_{\rm f, cho} = \frac{M}{\pi R_{\rm acc}^2 \rho_{\rm p} \delta v} =
  \frac{4}{18} \frac{R^2}{R_{\rm acc}^2} \frac{\rho_{\rm g}}{\rho_{\rm c}}
  t_{\rm f, gas} \, .
\end{equation}
We use similar expressions to equations
(\ref{eq:decc2dtdrag})-(\ref{eq:dinc2dtdrag}) for the damping by chondrule
accretion, but with $R^2$ replaced by $R_{\rm acc}^2$, $\rho_{\rm g}$ replaced
by $\rho_{\rm p}$, and multiplied by $18/4$ to take into account that the
pre-factor in the friction time of chondrule accretion ($4/3$) is much lower
than for gas drag ($6$). Damping by bouncing collisions and chondrule
scattering is taken into account by using the maximum of the Bondi radius and
the physical radius as the gravitational interaction radius when $t_{\rm f} >
t_{\rm B}$ and $R_{\rm acc}<R$.

\subsection{Planetesimal growth simulations}
 
The dynamical equations describing the temporal evolution of the masses,
eccentricities and inclinations of a large number of planetesimals are solved
in a numerical code named ``Pebble Accretion Onto Planetesimals and Planets''
(PAOPAP), which we have developed for the purpose of demonstrating the
importance of chondrule accretion for planetesimal growth. Information about
the simulations can be also found in the main paper; here we give some
additional details.

The planetesimals are treated as individual particles in the code, each
represented by a mass, an eccentricity and an inclination. Planetesimal masses
are evolved via the accretion of chondrules and mutual planetesimal collisions
(see Section \ref{s:collisions}), while the eccentricities and inclinations
are changed by both viscous stirring and dynamical friction from the other
planetesimals, as well as damping by gas drag and pebble accretion and
scattering. We ignore mass erosion of small ($<$ 50 km) planetesimals by large
chondrules impacting and bouncing at super-escape speeds.

The code divides the planetesimals into discrete size bins and calculates
$\dot{M}$, $\dot{e}$ and $\dot{i}$ for the smallest and largest planetesimal in
each bin. The temporal evolution of all other planetesimals is found by
interpolation from the two anchors in each bin. This approach allows us to
simulate a high number of planetesimals at a relatively low computational cost
and to resolve the run-away growth of a small number of large objects. We use
200 logarithmically spaced bins spanning from 10 km to 10,000 km in radius.

The chondrule density has a Gaussian stratification profile, with the
scale-height $H_{\rm p}$ in each bin set according to the
diffusion-sedimentation equilibrium expression
\begin{equation}
  \frac{H_{\rm c}}{H_{\rm g}} = \sqrt{\frac{\alpha}{\varOmega t_{\rm
  f} + \alpha}} \, . \label{eq:HcHg}
\end{equation}
Here $H_{\rm g}$ is the gas scale-height, $\alpha$ is the turbulent viscosity
(which we assume takes the same value as the turbulent diffusion coefficient),
$\varOmega$ is the Keplerian frequency at the orbital distance to the asteroid
belt and $t_{\rm f}$ is the friction time. The latter is given in the Epstein
regime by
\begin{equation}
  t_{\rm f}=\frac{a \rho_\bullet}{c_{\rm s}\rho_{\rm g}} \, ,
\end{equation}
with $a$ denoting the particle radius, $\rho_\bullet$ the material density,
$c_{\rm s}$ the sound speed in the gas (which depends only on the temperature
and the mean molecular weight), and $\rho_{\rm g}$ the gas density. We choose a
nominal turbulent diffusion coefficient of $\alpha=10^{-4}$ (we discuss this
choice in the main paper). This yields a scale-height of mm-sized particles of
approximately 30\% of the gas scale-height, with smaller particles having
scale-heights increasing as the square root of their size. 

We create chondrules continuously on a time-scale of 1.5 Myr to satisfy the
cosmochemical evidence that chondrules in individual chondrites have ages from
0 to 3 Myr\cite{Connelly+etal2012}. We have performed test simulations where
all the chondrules are present from the beginning; this accelerates chondrule
accretion but does not affect the overall picture that asteroids, Kuiper belt
objects and planetary embryos grow by chondrule accretion. The chondrules are
divided into 30 logarithmically spaced bins, with the number density $n(a)$
distributed according to the size distribution ${\rm d}n(a)/{\rm d} a \propto
a^{-3.5}$. 

The planetesimals have initial sizes from 10 to 150 km, distributed in a
shallow power law with the differential number following ${\rm d}N/{\rm d}R
\propto R^{-2.8}$, but truncated with an exponential term $\exp[-(R/R_{\rm
exp})^4]$ at the planetesimal radius $R_{\rm exp}$=100 km. The smallest
planetesimals experience very little growth and act mainly to provide dynamical
friction, damping the inclinations and eccentricities of the larger asteroids
and planetary embryos.

The pebble accretion rates of the planetesimals are found via interpolation in
a look-up table that gives the accretion radius for a grid of values of the
planetesimal size $R/R_{\rm B}$ (normalised by the Bondi radius) and the
chondrule friction time $t_{\rm f}/t_{\rm B}$ (normalised by the Bondi time).
The look-up table is based on a high number of integrations of the dynamics of
single chondrules passing by a planetesimal with the sub-Keplerian flow.

The temporal integration of the dynamical equations is done via a simple
Eulerian scheme, with the time-step determined to make $M$, $e$ and $i$ change
by a maximum of 10\%. We experimented with 30\% and found significant changes
to the results, while 3\% gives very similar results.

\subsection{Planetesimal collisions}
\label{s:collisions}

Planetesimal collisions are included in the code via a Monte Carlo method.
The planetesimal particles are first sorted in discrete size bins. For each bin
we calculate the average inclination and eccentricity of the planetesimals. We
then calculate the collision rate matrix for all combinations of bins,
$r_{ij}$. The collision rates are calculated following the scheme described in
the online supplement of Morbidelli et al.\ (2009)
\cite{Morbidelli+etal2009}. We refrain from describing the scheme in
detail here since it is already well described in Section 1.1 of the online
supplement of Morbidelli et al.\ (2009) \cite{Morbidelli+etal2009}.
\begin{figure}[!t]
  \begin{center}
    \includegraphics[width=0.45\linewidth]{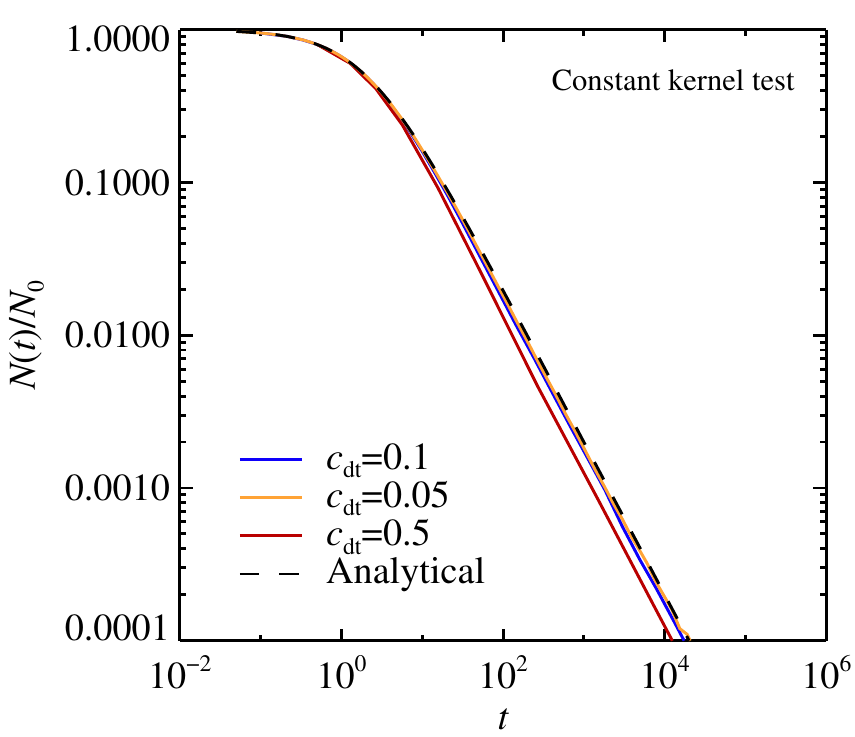}
  \end{center}
  \caption{The number of remaining bodies versus time for the constant kernel
  test. There is an excellent agreement between the results of the coagulation
  algorithm and the analytical solution for a time-step parameter below
  $c_{\rm dt}=0.5$. The time-step parameter is the ratio of the time-step to
  the collision time-scale.}
  \label{f:constant_kernel_test}
\end{figure}

From the time-step of the code we can then calculate the probability
$P_{ij}$ that a planetesimal from bin $i$ collides with any of the
planetesimals in bin $j$ during the time-step. If a random number $r$ is less
than $P_{ij}$ then the planetesimals will be collided. We assume perfect
sticking between planetesimals and add the mass of the smaller planetesimal to
the larger, removing afterwards the smaller planetesimal from the simulation.

We tested the coagulation scheme against an analytical solution to the
coagulation equation where the kernel is constant \cite{Wetherill1990}.
This corresponds to setting the collision rate $r_{ij}=K n_j$, where $K$ is a
constant and $n_j$ is the number density of planetesimals in bin $j$. The
solution to the constant kernel test is that the number of bodies remaining at
time $t$ is
\begin{equation}
  f = \frac{n(t)}{n_0} = \frac{1}{1+K n_0 t/2} \, .
\end{equation}
The number of remaining bodies versus time is shown in Figure
\ref{f:constant_kernel_test}, for different values of the time-step. The
time-step is set by
\begin{equation}
  {\rm d} t = c_{\rm dt} t_{\rm coll} \, ,
\end{equation}
where $t_{\rm coll}$ is the collision time-scale and $c_{\rm dt}$ is a
constant time-step parameter. There is excellent agreement in Figure
\ref{f:constant_kernel_test} between the analytical expression and the results
of the coagulation algorithm, for $c_{\rm dt}<0.5$. Passing this test implies
that the Monte Carlo collision detection scheme is implemented correctly.
\begin{figure}[!t]
  \begin{center}
    \includegraphics[width=0.45\linewidth]{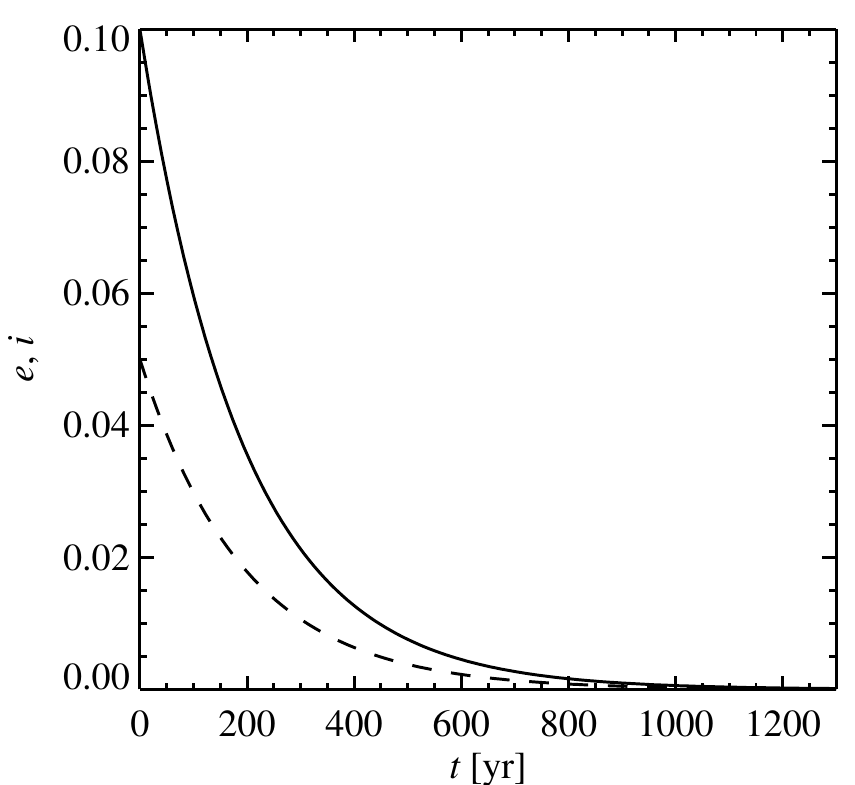}
  \end{center}
  \caption{Damping of eccentricities and inclinations of 10,000 1-cm-sized
  planetesimals located between 30 and 35 AU, emulating a test problem defined
  in Morbidelli et al.\ (2009) \cite{Morbidelli+etal2009}.  The total mass of
  the planetesimals is set to 10 Earth masses. The curves agree to within 15\%
  with those shown in Figure 12 of the supplemental material of Morbidelli et
  al.\ (2009).}
  \label{f:colldamp_test}
\end{figure}

A test of the collision rate calculation can not be done as simply, as the
expressions used in Morbidelli et al.\ (2009) \cite{Morbidelli+etal2009} are
complex. We choose therefore to reproduce one of the figures in that paper,
namely a plot of the damping of eccentricities and inclinations of
planetesimals due to mutual inelastic collisions. The damping is not very
relevant for the large planetesimals considered in this paper, but the exact
shape of the damping provides an excellent comparison with the collision rate
calculation of Morbidelli et al.\ (2009) \cite{Morbidelli+etal2009}. The test
problem considers 10 Earth masses of 1-cm-sized ``planetesimals'' located
between 30 and 35 AU. Collisions are assumed to be inelastic and lead to energy
dissipation; accretion and fragmentation are not included. Figure
\ref{f:colldamp_test} shows the evolution of the eccentricity (initially $0.1$)
and the inclination (initially $0.05$). There is agreement to within 15\%
between this plot and supplemental Figure 12 of Morbidelli et al.\ (2009)
\cite{Morbidelli+etal2009}. This shows that the collision rate algorithm is
likely implemented properly in the PAOPAP code.

\begin{figure}[!t]
  \begin{center}
    \includegraphics[width=0.45\linewidth]{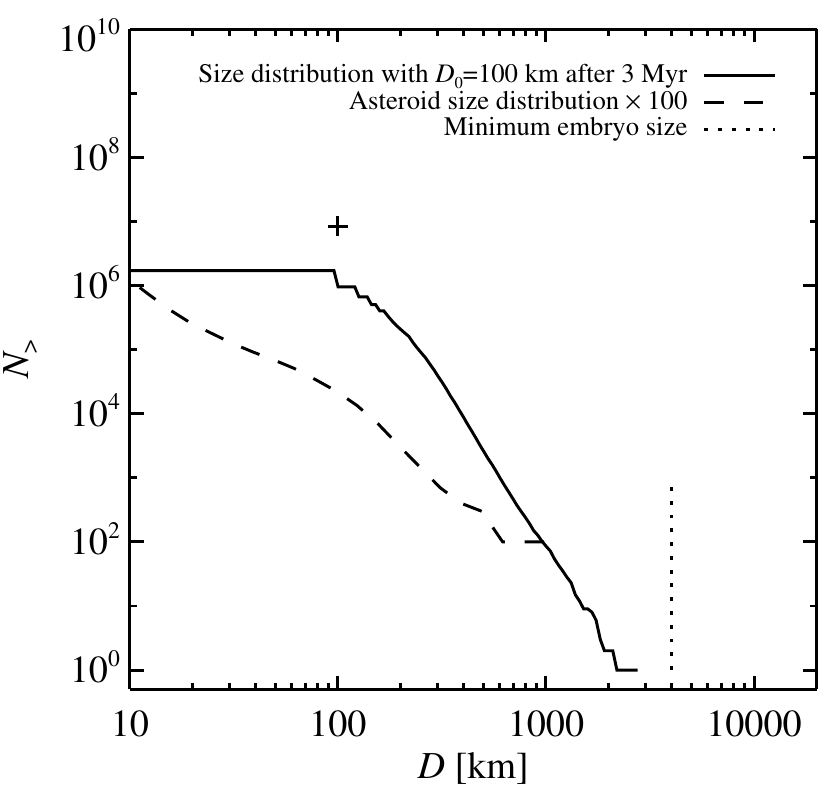}
  \end{center}
  \caption{Cumulative size distribution after 3 Myr of coagulation within a
  population of planetesimals of initial diameters 100 km (50 km in radius), as
  shown by the cross.  The size distribution is much steeper than the current
  observed size distribution in the asteroid belt. The results presented here
  are agree to within a few percent with Figure 5 of Morbidelli et al. (2009)
  \cite{Morbidelli+etal2009} and Figure 5 of Weidenschilling (2011)
  \cite{Weidenschilling2011}, except for $D>1000$ km where we have a slight
  underproduction of embryos. Note that we set the column density of
  planetesimals equal to that of solids in the Minimum Mass Solar Nebula of
  Hayashi (1981), while Weidenschilling (2011) uses approximately twice that
  value.}
  \label{f:Ncumu_D_R050km}
\end{figure}

An additional test was performed with actual planetesimals. The evolution of a
population of 50-km-radius planetesimals was presented in Figure 5 of
Morbidelli et al. (2009) \cite{Morbidelli+etal2009} and Figure 5 of
Weidenschilling (2011) \cite{Weidenschilling2011}. We performed a similar
simulation with our planetesimal accretion code. The resulting cumulative size
distribution is shown in Figure \ref{f:Ncumu_D_R050km}. The cumulative size
distribution is very steep and terminates just before reaching embryo sizes.
This steep cumulative size distribution agrees to within a few percent with the
results of Morbidelli et al. (2009) \cite{Morbidelli+etal2009} and
Weidenschilling (2011) \cite{Weidenschilling2011}, except for a slightly
smaller production of the very largest embryos in our simulations. Note that
these authors include fragmentation of planetesimals in their calculations,
while we ignore this effect. The agreement of the results for $D>20$ km shows
that planetesimal fragmentation plays little role for the coagulation of such
large planetesimals. It is well-known that planetesimals larger than 100 km are
strong and can survive high collision speeds \cite{Bottke+etal2005}.

\clearpage

\section{Supplementary Text}

\subsection{Planetesimal formation in the asteroid belt}

The particle sizes used in our streaming instability simulations correspond to
approximately 25-cm-sized rocks at 2.5 AU in the Minimum Mass Solar Nebula
(Stokes number ${\rm St}=0.3$).  This is much larger than the typical diameters
of chondrules. We have chosen to study the formation of planetesimals from such
large particle sizes in order to have a clean convergence test in parameters
employed in previous works \cite{Johansen+etal2009,Johansen+etal2012}. We
suggest three ways to understand the actual formation of planetesimals in the
asteroid belt:

{\it 1) Chondrule aggregates.} Chondrules are covered with rims of
fine-grained matrix \cite{Metzler+etal1992} and these rims can facilitate
the formation of chondrule aggregates. If the turbulent speed of the gas is
very low, with turbulent viscosity $\alpha\sim10^{-6}$, then collision speeds
between rimmed chondrules are low enough to allow chondrules to stick together
\cite{Ormel+etal2008}. The porous rims act as shock absorbers and
facilitate the formation of decimeter-sized chondrule aggregates. These
aggregates are large enough to trigger streaming instabilities and the direct
formation of planetesimal seeds that can go on to accrete individual
chondrules, as envisioned in the main paper. The radial drift speed of these
aggregates is high, approximately 10 m/s, causing the aggregates to drift
inwards within a few thousand years \cite{Weidenschilling1977}. The
combination of growth and drift provides a means (a) to form planetesimals from
such aggregates and (b) to subsequently flush these aggregates towards the
inner solar system, to avoid feeding chondrule aggregates to the largest
planetesimals, as this would trigger a very efficient run-away growth of
asteroids and give a too steep size distribution.

\begin{figure*}
  \begin{center}
    \includegraphics[width=0.8\linewidth]{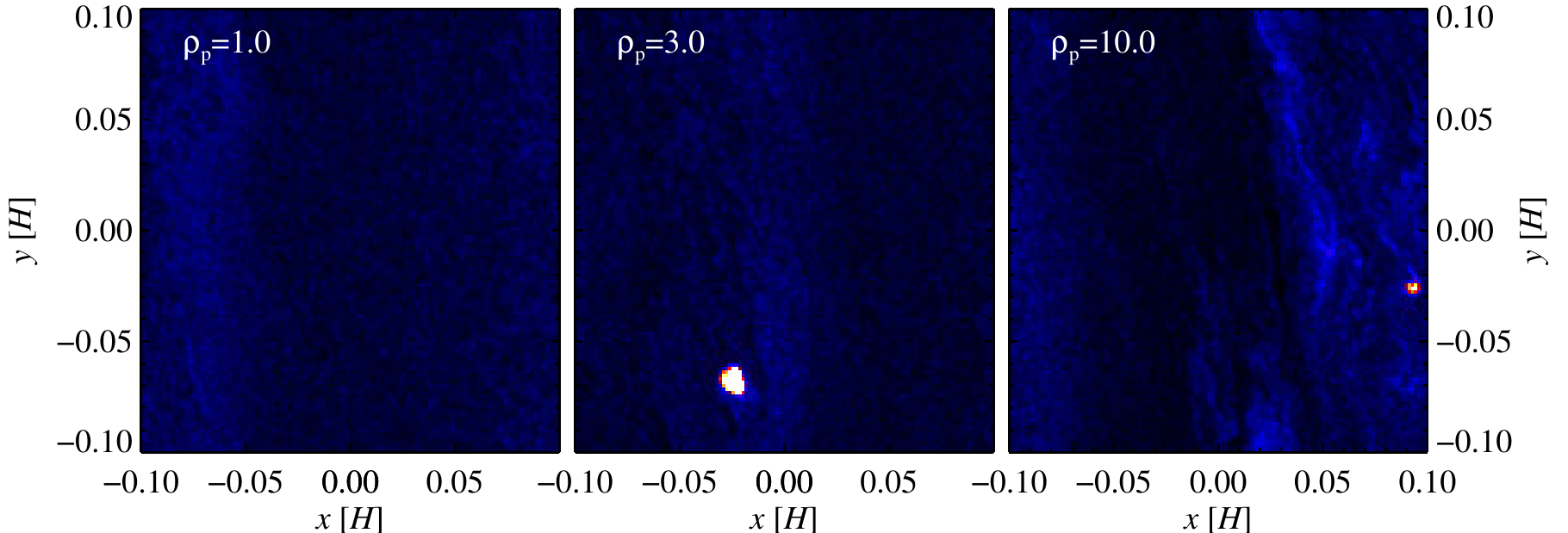}
  \end{center}
  \caption{Formation of planetesimals from centimeter-sized particles in a
  2-D shearing sheet simulation with the same spatial extent as the streaming
  instability simulations presented in the main text. The mean particle density
  is $\rho_{\rm p}=1$ in the left panel, $\rho_{\rm p}=3$ in the middle panel
  and $\rho_{\rm p}=10$ in the right panel (relative to the gas density). The
  mean density is set to mimic the effect of both stratification and filament
  formation by the streaming instability.  The simulations with $\rho_{\rm
  p}=3$ and $\rho_{\rm p}=10$ form a single, large planetesimal of
  approximately 100 km in radius after a few hundred years of integration. This
  numerical experiment shows that planetesimals can form for particle sizes
  corresponding to centimeter-sized macrochondrules and that planetesimal sizes
  are similar to those forming out of 25-cm-sized particles presented in Figure
  \ref{f:size_distribution} of the main text.}
  \label{f:planetesimals_cm_particles}
\end{figure*}

{\it 2) Large chondrules.} Another possibility is that the first planetesimal
seeds formed out of centimeter-sized macrochondrules
\cite{WeyrauchBischoff2012}. These could have drifted out of the asteroid belt
subsequent to planetesimal formation and left the stage for accretion of their
millimeter-sized counterparts. Carrera, Johansen, \& Davies
\cite{Carrera+etal2014} investigate the ability for the streaming instability
to form dense filaments in particles down to chondrule sizes. While
chondrule-sized particles are very hard to concentrate, particles just a few
times larger (Stokes number of 0.01) concentrate readily into filaments at a
particle mass loading of a few times the nominal value in the solar
protoplanetary disk ($Z_\odot \sim 0.01$). Here we explore whether these
conditions are susceptible to planetesimal formation, by including the
self-gravity between centimeter-sized particles (Stokes number of 0.01, so 30
times smaller than in the simulations presented in Figure
\ref{f:size_distribution} of the main paper). We use 2-D shearing sheet
simulations and set the mean particle density to 1, 3 and 10 times the gas
density, respectively, to mimic the mid-plane conditions seen in Carrera,
Johansen, \& Davies \cite{Carrera+etal2014}. The results are shown in Figure
\ref{f:planetesimals_cm_particles}. The simulations with particle density
$\rho_{\rm p}=3\rho_{\rm g}$ and $\rho_{\rm p}=10\rho_{\rm g}$ form a single
planetesimal after a few hundred years. These planetesimals have a
characteristic radius of approximately 100 km. This shows that planetesimals
can indeed form out of particles as small as 1 cm in size and that the
resulting planetesimal sizes are similar to those seen in Figure
\ref{f:size_distribution} of the main paper. Thus we show that asteroids can
form from centimeter-sized particles.  The full exploration of the relevant
parameters for formation of planetesimals from centimeter-sized particles is
beyond the scope of this paper, but this should have a high priority in future
research.

{\it 3) Icy pebbles.} It is possible that the ice line in the solar
protoplanetary disk was much closer to the star in the earliest phases of
protoplanetary disk evolution.  Hence the asteroid belt could have been
populated originally with icy pebbles of sizes comparable to what we used in
the simulations\cite{RosJohansen2013}. The first generation of icy asteroid
seeds could subsequently have dried out after heating by $^{26}$Al or simply
have been swamped in the subsequent accretion of dry chondrules.  Carbonaceous
chondrites display a wide range in their degree of aqueous alteration,
consistent with the accretion and subsequent melting of icy pebbles
\cite{Krot+etal2003}. However, even some ordinary chondrites have been
shown to have experienced water flows \cite{Hutchison1987}; and hence the
formation of the seeds of the nowadays dry ordinary chondrites could have
involved accretion of icy material.

Irrespective of the details of asteroid formation, the important factors
for our chondrule accretion model are (i) planetesimals form, (ii) the
planetesimals are not too small to prevent significant chondrule accretion, and
(iii) planetesimals are not so large that all chondrules are accreted on the
largest objects in the population. The computer simulations of planetesimal
formation by streaming instabilities inspire us to take asteroid seeds with
characteristic radii of 100 km and with a fairly shallow size distribution
that puts most of the mass in the largest bodies.

\subsection{Particle growth and chondrule precursors}

In our model we assume that the centimeter-to-decimeter-sized particles
discussed in the previous section were only present during the earliest stages
of planet formation when the planetesimals formed. Ice particles could only
exist while the ice line was interior of the asteroid belt, while the formation
of chondrule aggregates requires very weak turbulence of $\alpha\sim10^{-6}$
\cite{Ormel+etal2008}, which may have required an extensive dead zone in a
young (and hence still massive) protoplanetary disk. In this section we discuss
why the later stages of the protoplanetary disk were likely dominated by
chondrule-sized particles.

The growth of dust particles in protoplanetary disks happens initially through
sticking collisions that lead to the formation of fluffy dust aggregates; see
the review paper on coagulation and planetesimal formation by Johansen et al.\
\cite{Johansen+etal2014}. These aggregates are compactified by mutual
collisions as they reach approximately millimeter sizes.  Compact dust
aggregates have poor sticking properties and bounce off each other when they
collide \cite{Guettler+etal2010}. This {\it bouncing barrier} is a major
obstacle for dust growth \cite{Zsom+etal2010}, which could otherwise continue
to much larger sizes \cite{Weidenschilling1997,Weidenschilling2000}. Particles
can still grow by mass transfer in high-speed collisions; however the growth
rates remain too low to compete against radial drift \cite{Windmark+etal2012}.
Another possibility is that ice particles could be extremely fluffy and
experience perfect sticking even at high speeds \cite{Okuzumi+etal2012}.  This
nevertheless requires aggregates to consist of very small monomers (0.1
$\mu$m), while matrix in chondrites is typically at least 10 times larger.

Hence it appears that the growth of silicate dust grains in the asteroid
formation region stops at around mm sizes. This bouncing barrier picture fits
well with the sizes of chondrules found in chondrite meteorites. The actual
formation process of chondrules is highly uncertain, but is believed to involve
heating of dust aggregate precursors in shocks
\cite{DeschConnolly2002,Ciesla+etal2004}, in current sheets
\cite{McNally+etal2013} in the protoplanetary disk or in impacts between molten
planetesimals \cite{SandersTaylor2005}. Our model does not include the presence
of chondrule precursors. Such precursors could have a low fractal dimension and
hence low friction time, which would prevent sedimentation and efficient
incorporation into the asteroids and embryos that grow in the mid-plane.
Nevertheless the inclusion of chondrule precursors would not change the
conclusions of our paper, as they would simply lead to a slightly higher
accretion rate for asteroids and embryos.

\subsection{Depletion of the asteroid belt}

The current mass of the asteroid belt is only approximately $5 \times 10^{-4}$
Earth masses. This contrasts with the extrapolation of the Minimum Mass Solar
Nebula into the asteroid belt, which yields a particle column density of $4.3$
g/cm$^{2}$ and hence a primordial mass of 0.86 Earth masses. The depletion of
the asteroid belt has happened mainly through collisional grinding and pumping
of the eccentricity of asteroids in resonances with Jupiter. Based on a number
of arguments, including the extant number of asteroid families and the number
of large craters on Vesta, Bottke et al.\ \cite{Bottke+etal2005} concluded that
the collisional activity in the asteroid belt, after excitation of the
asteroids to their current eccentricities, can not have been much larger than
the current collision rate over 10 Gyr. Hence the asteroid belt must have been
depleted to near its current mass very soon after asteroid formation.
Morbidelli et al.\ \cite{Morbidelli+etal2009} calculate that the asteroid belt
could have lost only 1/3 of its current mass over a 10 Gyr equivalent of its
present collisional activity. This low level of collisional grinding implies
that the current size distribution of asteroids larger than 60 km in radius is
primordial \cite{Bottke+etal2005}.

The formation of embryos in the asteroid belt provides a means to excite the
smaller asteroids to their current, high eccentricities
\cite{Wetherill1992}. Embryos can also be responsible for the rapid
clearing of the asteroid belt, since asteroids are gravitationally scattered by
the embryos and hence move readily to resonances with Jupiter from which they
are removed from the belt \cite{ChambersWetherill2001}. Hence the formation
of planetary embryos in our chondrule accretion model is completely consistent
with mechanisms for depletion of the asteroid belt.

The main alternatives to the embryo model for asteroid belt depletion are (i)
sweeping resonances \cite{MintonMalhotra2009} and (ii) the migration of Jupiter
in the Grand Tack model \cite{Walsh+etal2011}. The migration of Jupiter and
Saturn during the restructuring of the giant planets to their current orbits by
scattering of the primordial Kuiper belt leads to a sweeping of resonant
locations across the asteroid belt. This could have led to the loss of 90-95\%
of the mass of the asteroid belt during the migration period
\cite{MintonMalhotra2009}. The Grand Tack model provides a more violent means
for Jupiter's gravity to interfere with the asteroid belt, as Jupiter migrates
through the asteroid belt twice; the change from inwards to outwards migration
is caused by the formation of Saturn, which comes to share a gap with Jupiter
in the protoplanetary disk. This injects C-type asteroids from beyond the ice
line into the asteroid belt, which ends up strongly depleted; the later
planetesimal-driven migration could inject further icy planetesimals into the
main belt \cite{Levison+etal2009}.

Our simulations show that embryos form readily in the asteroid belt from the
accretion of chondrules onto planetesimals. These embryos could have played an
important role in sculpting the current orbits of the asteroids and depleting
the asteroid belt. There are, as discussed above, mechanisms for the depletion
of the asteroid belt that do not require the presence of embryos there. The
reality could be that a combination of embedded embryos, sweeping resonances
and Jupiter's migration conspired to deplete the asteroid belt to its current
mass.


\end{document}